\documentclass[preprint]{iucr}
\RequirePackage{graphicx}  
\usepackage{amsmath}
\usepackage{amssymb}

\paperprodcode{a000000} 
\paperref{xx9999} 
\papertype{TR} 
\paperlang{english} 
\journalcode{B} 
\journalyr{2015}
\journaliss{}
\journalvol{}
\journalfirstpage{000}
\journallastpage{000}
\journalreceived{0 XXXXXXX 0000}
\journalaccepted{0 XXXXXXX 0000}
\journalonline{0 XXXXXXX 0000}

\newcommand{\mct}[1]{\multicolumn{2}{c}{#1}}
\newcommand{\mcnb}[2]{\multicolumn{#1}{c|}{#2}}

\begin{document}

\title{Hexagonal \emph{R}MnO$_3$: a model system for 2D triangular lattice antiferromagnets}

\shorttitle{Hexagonal \emph{R}MnO$_3$}

\author{Hasung}{Sim}
\author{Joosung}{Oh}
\author{Jaehong}{Jeong}
\author{Manh Duc}{Le}
\cauthor{Je-Geun}{Park}{jgpark10@snu.ac.kr}{}

\aff{Center for Correlated Electron Systems, Institute for Basic Science (IBS) \& Department of Physics and Astronomy, Seoul National University, \city{Seoul 08826}, \country{Korea}}

\shortauthor{Hasung Sim \emph{et al.}}

\maketitle

\begin{synopsis}
Supply a synopsis of the paper for inclusion in the Table of Contents.
\end{synopsis}

\begin{abstract}
The hexagonal \emph{R}MnO$_3$ (\emph{h}-\emph{R}MnO$_3$) are multiferroic materials, which exhibit the coexistence of a magnetic order and ferroelectricity. Their distinction is in their geometry that both results in an unusual mechanism to break inversion symmetry, and also produces a 2D triangular lattice of Mn spins, which is subject to geometrical magnetic frustration due to the antiferromagnetic interactions between nearest neighbour Mn ions. This unique combination makes the \emph{h}-\emph{R}MnO$_3$ a model system to test ideas of spin-lattice coupling, particularly when the both the improper ferroelectricity and the Mn trimerisation that appears to determine the symmetry of the magnetic structure arise from the same structure distortion. In this review, we demonstrate how the use of both neutron and X-ray diffraction and inelastic neutron scattering techniques have been essential to paint this comprehensive and coherent picture of the \emph{h}-\emph{R}MnO$_3$.
\end{abstract}

\section{Introduction}

The magnetic and crystal structures of multiferroic materials play a crucial role in determining their physical and functional properties. In the case of some of the perovskite manganites, it was established that ferroelectric order follows as a result of a spiral magnetic structure and the inverse Dzyaloshinskii-Moriya interaction~\cite{Kimura2003}. In others, a zig-zag magnetic order gives rise to ionic displacements via exchangestriction~\cite{Mochizuki2011}.
Similar mechanisms may lie at the root of the magnetoelectric coupling in the hexagonal manganite family, which is the focus of this article. Whilst \emph{R}MnO$_3$ compounds with lighter rare-earth elements (\emph{R}=La--Ho) can be stabilised with an orthorhombic-distorted perovskite structure, the smaller ionic radii of elements at the end of the lanthanide series results in a close-packed hexagonal structure with space group $P6_3/mmc$ at high temperatures.

Unlike the perovskite manganites, where magnetic and ferroelectric ordering temperatures coincide, the hexagonal manganites are \emph{type-I} multiferroics with quite different transition temperatures: ferroelectric $T_\mathrm{C}$~($>$1000~K) and magnetic $T_\mathrm{N}$~($<$100~K). This is because inversion symmetry is broken in these materials by the cooperative rotation of MnO$_5$ bipyramids~\cite{VanAken2004a} rather than due to a noncentrosymmetric magnetic structure. Nonetheless, strong magnetoelastic coupling effects have been observed in the hexagonal manganites, notably a large displacement of Mn ions further towards or away from their apical oxygen ion at the N\'eel temperature~\cite{Lee2008}. The initial Mn off-centering, however, occurs at the ferroelectric Curie point, and appears to correlate with the rare-earth ionic size. In addition to being exaggerated by the magnetic ordering, the nature of the initial Mn off-centering (whether towards or away from the apical oxygen) appears to determine the symmetry of the magnetic structure~\cite{Fabreges2009}. 

There is thus a strong link between the magnetic and crystal structures, and this carries over into the crystal and magnetic dynamics. For example, it has been recognised that there is a large degree of coupling between the magnons and phonons in the hexagonal manganites~\cite{Oh2015}, although no electromagnons have yet been reported. Finally, the magnon spectrum has also yielded evidence of magnon decay and nonlinear magnon-magnon interactions in a relatively large spin ($S=2$) system, due to the noncollinear magnetic structure arising from the geometrically frustrated triangular lattice of Mn$^{3+}$ ions~\cite{Oh2013}. In this article we will review and explore both aspects of the magneto-electric coupling, with the structural aspects discussed in section~\ref{sec-struct} and the dynamical properties in section~\ref{sec-excit}.

A short note of disclaimer: Although we tried to be comprehensive in covering the physics of \emph{h}-\emph{R}MnO$_3$, inevitably we could not include all the interesting topics of \emph{h}-\emph{R}MnO$_3$ in our article. Mainly because of the lack of the space, here we focused on the spin-lattice issue in a bulk form, leaving out some other interesting works and different properties in a nanocrystalline~\cite{Bergum2011} or film form, yet less related to our main point. 

\section{Structure} \label{sec-struct}

\begin{figure} \label{fig1}
\caption{Summary of phase diagrams of several hexagonal \emph{R}MnO$_3$. Transition temperatures are taken after \citeasnoun{Chae2012}, \citeasnoun{Lonkai2004}, \citeasnoun{Abrahams2001}, \citeasnoun{Gibbs2011}, \citeasnoun{Fan2014} and \citeasnoun{Lorenz2013}.}
\scalebox{.6}{\includegraphics{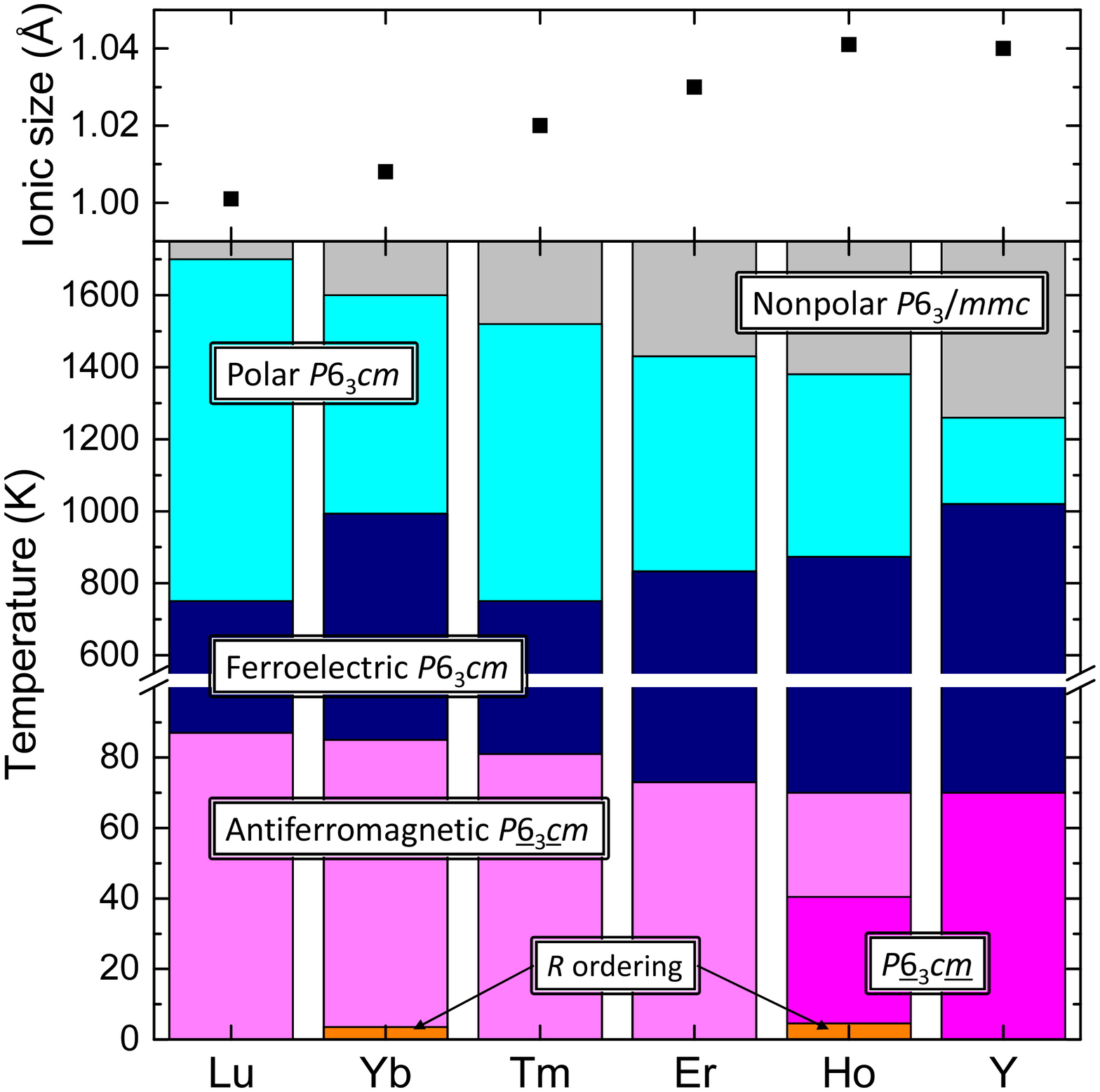}}
\end{figure}

The rare-earth manganite, \emph{R}MnO$_3$, compounds adopts one of two polymorphs: a distorted perovskite structure which is stabilised for larger $R^{3+}$ cations; and a hexagonal polymorph which is a stable phase for smaller $R^{3+}$. For intermediate sized cations, either structures may be stabilised by growth in oxygen-excess or -deficient atmosphere~\cite{Harikrishnan2009} or with the application of pressure~\cite{Zhou2006}. Whilst they exhibit both ferroelectricity and antiferromagnetism, the magnetoelectric coupling between them seems likely to occur via distortions of the crystal structure. The ferroelectric Curie temperature is around $\gtrsim$1000~K and has a slight dependence on the cation size, with YMnO$_3$ having the lowest $T_\mathrm{C}$ and the largest ionic radius. The N\'eel temperature is some ten times lower, $T_\mathrm{N}\lesssim100$~K, which may be due to the geometrical magnetic frustration of the triangular lattice of Mn spins. We note that the superexchange interactions between nearest neighbour Mn-Mn pairs is quite strong, giving a Curie-Weiss temperature (which is proportional to the sum of the exchange interactions) of $\approx$$-$600~K. The transition temperatures and the crystal and magnetic space group symmetry is summarized in figure~\ref{fig1}, in the order of decreasing $R^{3+}$ cation size. 
As we noted above, the actual magnetic ordering is pushed towards much lower temperature probably because of the intrinsic geometrical frustration of the triangular lattice and also the low dimensionality. Therefore, we do not think that the big difference between the FE and AFM transition temperatures itself indicates a weaker magnetoelectric coupling for $h$-$R$MnO$_3$ although this argument has been used in some corner of the community.

Whilst some studies have reported only a single phase transition above room temperatures, others have found two, which has led to divergent views on the nature of the ferroelectric transition and the origin of ferroelectricity in the hexagonal manganites. There are two principle structure distortions that lowers the symmetry of the system from non-polar (paraelectric) $P6_3/mmc$ to polar (ferroelectric) $P6_3cm$. Whilst the $\Gamma_2^-$ mode produces a net polarisation, the unit cell tripling $K_3$ mode does not.
Calculations show, however, that the $K_3$ mode is the primary order parameter that induces the $\Gamma_2^-$ distortion due to geometric factors~\cite{VanAken2004a}, making the hexagonal manganites improper ferroelectrics. As the $K_3$ mode also results in the trimerisation of the Mn sublattice, it affects, and is affected by, the magnetic ordering and so provides a microscopic mechanism for the magneto-electric coupling. If, on the other hand, the two distortions are independent as may be the case if two distinct transitions exist at which each distortion is stabilised, then this mechanism is invalid. 

We thus begin this section with a discussion of the high temperature transitions, and the nature of the ferroelectricity, before moving on to a discussion of the magnetic structure and its connection to the crystal structure and trimerisation distortion.

\subsection{The ferroelectric transition} \label{subsec-fe}

\begin{figure} \label{fig2}
\caption{Analysis of symmetry and structural changes at high temperature. Four possible routes from $P6_3/mmc$ to $P6_3cm$ that are related to the ferroelectric and structural transitions. The symmetry analysis are adopted from \citeasnoun{Lonkai2004}, \citeasnoun{Nenert2005} and \citeasnoun{Fennie2005}.}
\scalebox{.6}{\includegraphics{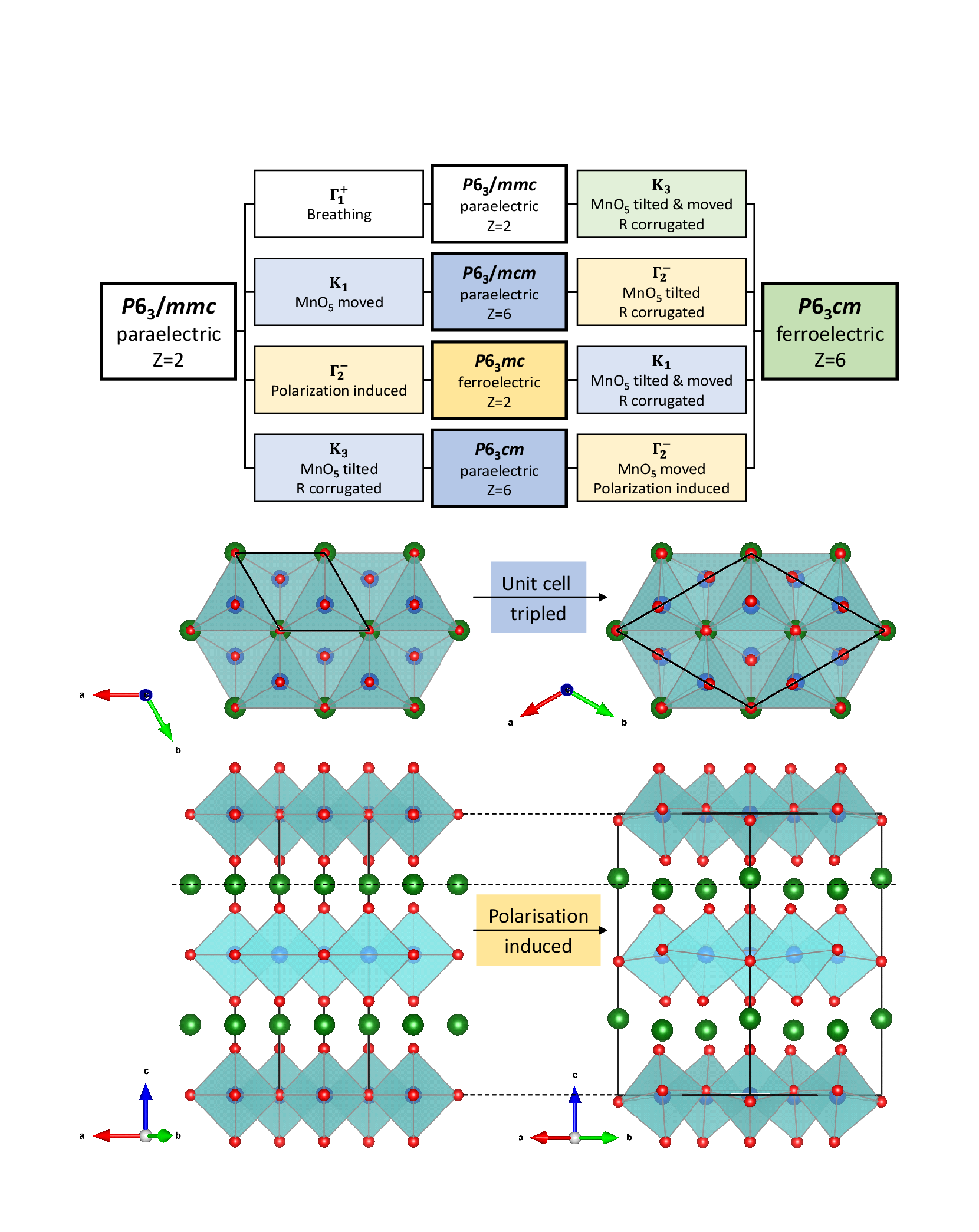}}
\end{figure}

Figure~\ref{fig2} shows the crystal structures of the non-polar (paraelectric) $P6_3/mmc$ and polar (ferroelectric) $P6_3cm$ phases. The four space groups that are both subgroups of $P6_3/mmc$ and supergroups of $P6_3cm$ are each associated with a symmetry lowering mode~\cite{Lonkai2004}, and their relationship is also shown in figure~\ref{fig2}. The $\Gamma_1^+$ breathing mode affects only the $z$ position of the apical oxygen and does not change the space group symmetry. The two $K$ modes result in a $\sqrt{3}\times\sqrt{3}$ tripling of the unit cell, either by tilting of the MnO$_5$ trigonal bipyramid ($K_3$) or its displacement along $a$ ($K_1$). This results in extra peaks in the diffraction pattern that are clearly visible in the experimental data. However, these modes do not produce a net ferroelectric polarisation, although the $K_3$ mode produces a local dipole moment, this is cancelled globally. Rather, the ferroelectricity only arises from the $\Gamma_2^-$ distortion, which allows the displacements of the $R$ and Mn cations and oxygen anions with respects to each other along the $c$ axis. However, this distortion by itself does not result in a unit cell tripling and can yield a proper ferroelectric phase with $P6_3mc$ symmetry. 

Early dielectric constants~\cite{Coeure1966} and pyroelectric current~\cite{Ismailzade1965} measurements suggested that the ferroelectric transition should be below 1000~K, which is correlated with a change in the slope of the resistivity~\cite{Choi2010}. However, neutron~\cite{Lonkai2004,Gibbs2011} and X-ray~\cite{Lonkai2004,Nenert2007} diffraction studies indicated a unit cell tripling at higher temperatures $\approx$1250~K. These observations can only be reconciled if the higher temperature transition arises either from the $K_1$ or $K_3$ mode, whilst the $\Gamma_2^-$ mode is stabilised below the lower temperature transition. This would yield either a paraelectric $P6_3/mcm$ or antiferroelectric $P6_3cm$ intermediate phase. The former case was favoured by \citeasnoun{Nenert2005}, whilst \citeasnoun{Lonkai2004} and \citeasnoun{Gibbs2011} showed from detailed analysis of their neutron diffraction patterns that the MnO$_5$ bipyramid is indeed tilted rather than simply displaced, establishing that the $K_3$ mode is stabilised and the intermediate structure is $P6_3cm$. 

This scenario is further supported by \emph{ab initio} calculations, which showed that the $K_3$ mode is strongly unstable in the symmetric $P6_3/mmc$ structure~\cite{Fennie2005}, whereas the $K_1$ mode is stable with high calculated phonon frequencies. A decomposition of the atomic displacements between the $P6_3/mmc$ structure and the room temperature $P6_3cm$ structure in terms of the normal modes also shows that the amplitude of the $K_3$ mode (0.93\AA) is much greater than the $K_1$ (0.03\AA), or $\Gamma_2^-$ modes (0.16\AA). 

Considering all the experimental and theoretical studies together, it is of our view that the first high temperature transition above 1200~K is from $P6_3/mcm$ to $P6_3cm$ while the second transition at around 900--1000~K is the isostructural transition involving a huge increase of electric polarization and so the intermediate phase is the polar $P6_3cm$ space group. Because of this polar nature of the intermediate phase, it is most likely that $h$-$R$MnO$_3$ already has nonzero electric polarization below the first high temperature phase transition although it seems to have a smaller value. Only when it undergoes the second isostructural transition below 1000~K, it begins to develop the large polarization value of around 5~$\mu$C/cm$^2$ at room temperature.

\subsubsection{Origin of ferroelectricity} \label{subsec-fe-origin}

The first principles calculations point to a mechanism underlying the ferroelectricity in the $h$-$R$MnO$_3$ system. \citeasnoun{VanAken2004} were the first to suggest the principles of what was later termed as ``geometric ferroelectricity", in which in certain geometries global inversion symmetry may be broken by a polyhedral tilt. For the $h$-$R$MnO$_3$, the triangular symmetry of the Mn-O plane means that the $K_3$ tilt of the MnO$_5$ bipyramid satisfies this condition, which is not the case for the octahedral tilts of the perovskite structure. The next essential ingredient is the coupling of this distortion to the polar mode $\Gamma_2^-$, which \citeasnoun{Fennie2005} showed to have a nonzero equilibrium displacement when the amplitude of the $K_3$ mode is finite. Thus the $K_3$ mode acts as a ``geometric field" that pushes the equatorial oxygen ions away from the Mn plane, giving unequal $R$-O$_{\mathrm{eq}}$ distances due to the buckling of the $R$-layer, which accompanies the MnO$_5$ tilt.

\begin{figure} \label{fig3}
\caption{(Top) Simulated diffraction patterns for four possible space groups as shown in Fig.~\ref{fig2}. A region of interests in the diffraction patterns is marked by shading. (Bottom) Temperature dependence of our high-resolution x-ray diffraction patterns for LuMnO$_3$ and YMnO$_3$ taken at high temperature.}
\scalebox{.7}{\includegraphics{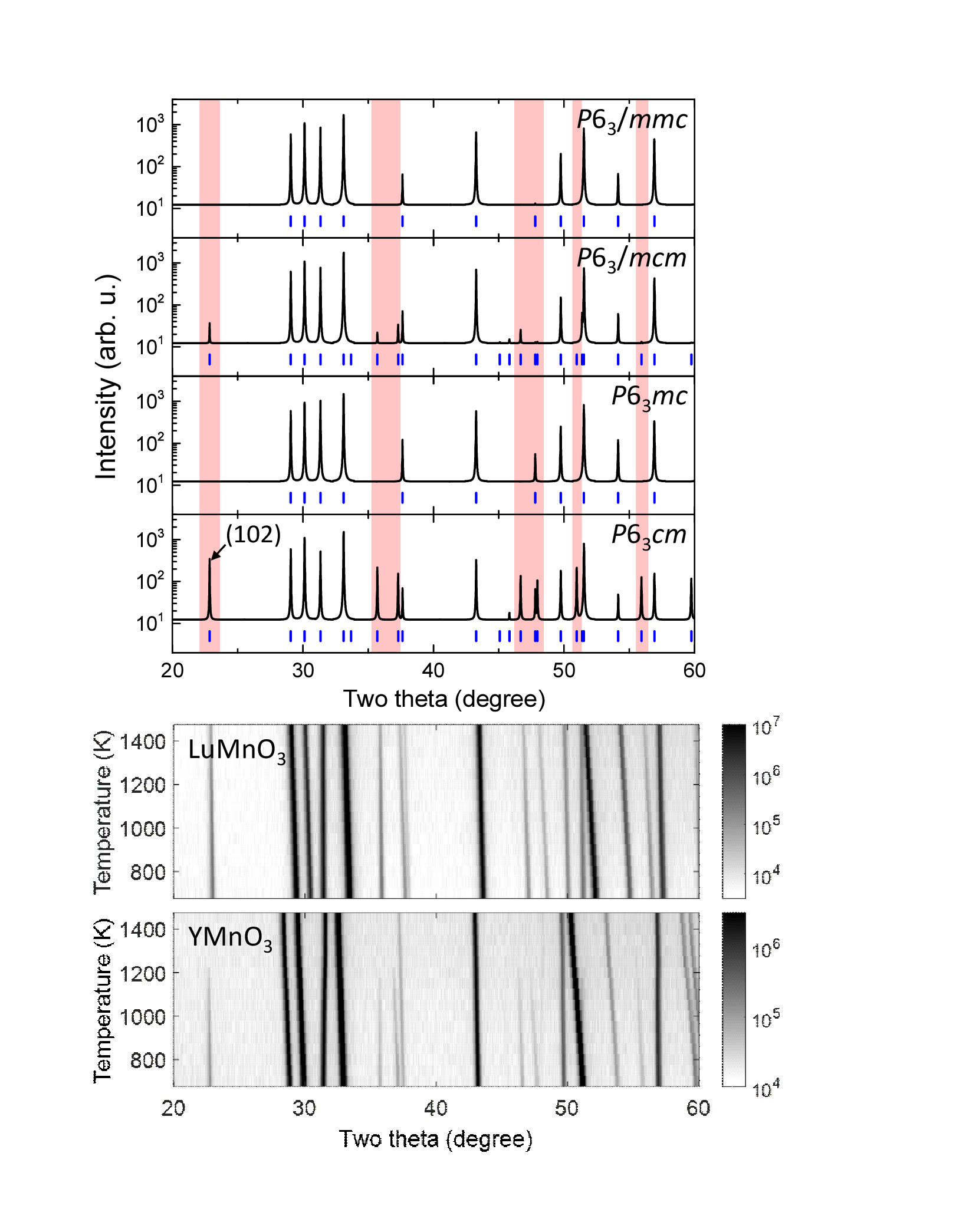}}
\end{figure}

\begin{figure} \label{fig4}
\caption{(Top) Temperature dependence of the integrated intensity of the 102 Bragg peak and the lattice parameters in YMnO$_3$. The lines represent our theoretical calculations using Landau-Ginzburg analysis with one order parameter (solid) and two order parameters (dashed line). The structural transition ($P6_3/mmc$ to $P6_3cm$) and the secondary transition are clearly visible at 1225 and 1012~K, respectively. (Bottom) The temperature dependence is presented of the two lattice constants with the lines serving as guides for eye at both below and above the second transition at 1012~K.}
\scalebox{.6}{\includegraphics{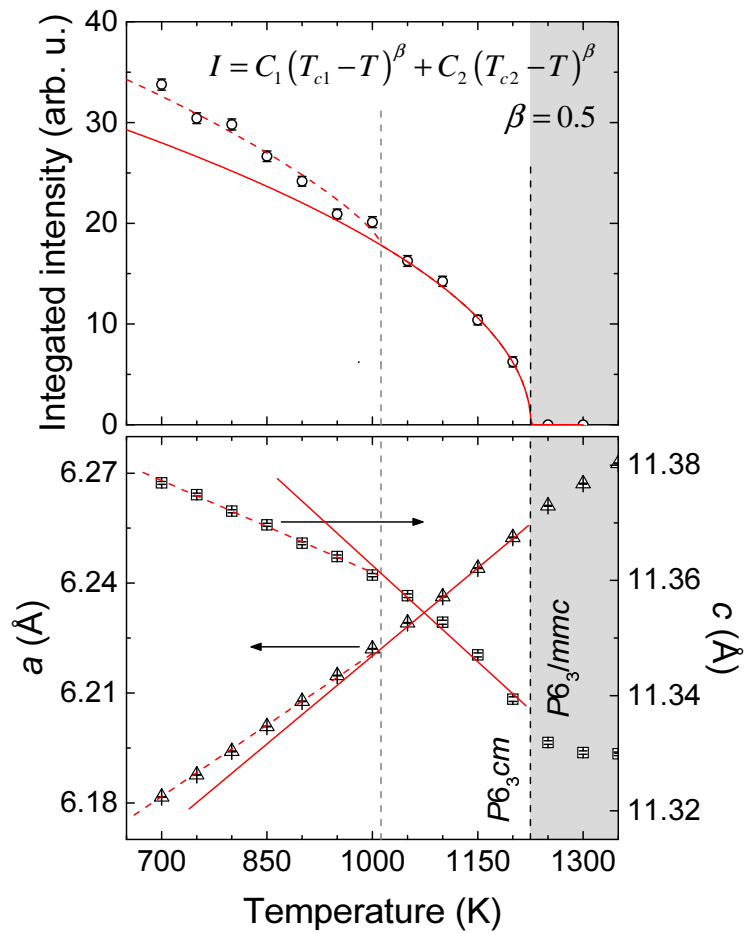}}
\end{figure}

Although this coupling is initially nonlinear and small, it only becomes linear and significant above a cross-over threshold. This cross-over temperature is calculated to be $\approx$100~K~\cite{Fennie2005}, which is about the same order as the difference between the upper and lower transition temperatures seen in the diffraction and physical properties measurements as discussed above. Thus, the two transitions may be explained, in part, by the nature of the ferroelectricity in the $h$-$R$MnO$_3$; although a finite polarisation exists below the initial structure transition between $P6_3/mmc$ and $P6_3cm$, it only becomes significant after ``turning on" the polar mode of $\Gamma_2^-$ at a lower temperature. This scenario may be supported by our high resolution X-ray diffraction measurements at high temperatures, shown in figure~\ref{fig3}. Peaks from the tripled unit cell, outlined in red in figure~\ref{fig3}, appear below $\approx$1250~K, which correlates well with a sharp increase of the $c$ lattice constant, shown in figure~\ref{fig4}. 
The temperature dependence of the integrated intensity of 102 Bragg peaks, drawn in figure~\ref{fig4}, is best fit by a model with two transitions at 1225(9) and 1012(32)~K, if the critical exponent is restricted to be $\beta=\frac{1}{2}$ required for a second order Landau phase transition. The ratio of the magnitude of the upper to lower transitions, 4.66, is also close to the amplitude ratio of the $K_3$ and $\Gamma_2^-$ modes, 5.8 as found in the theoretical studies~\cite{Fennie2005}, suggesting that the upper transition may be due to the $K_3$ mode and the lower transition to the $\Gamma_2^-$ mode.

As another indicator for the source of the ferroelectricity, the Born effective charges estimated from the first principles calculations by \citeasnoun{VanAken2004} were found to be quite close to the nominal valences, indicating that the ferroelectricity should not result from strong hybridisation effects. However, \citeasnoun{Cho2007} observed several peaks in the oxygen K-edge X-ray absorption spectrum, which may only be explained by a strong overlap between the empty $d$-states of rare-earth elements and the O $p$-states. Further, the measurements also showed striking differences depending on whether the incident light was polarised parallel or perpendicular to the $c$ axis, indicating that this hybridisation is highly anisotropic and stronger along the $c$ axis. This is consistent with latter optical conductivity measurements by~\citeasnoun{Zaghrioui2008}, who determined that the Born effective charge tensor is anisotropic with $Z_{zz}^*(\mathrm{O})\sim-3$ and $Z_{zz}^*(R,\mathrm{Mn})\sim4.5$, relatively enhanced compared to the ionic expectations. Similarly a separate X-ray diffraction study using the maximum entropy method (MEM) by \citeasnoun{Kim2009} showed an increased hybridisation effect between $R$ ions at the $2a$ Wyckoff sites and the equilateral O ions below the second transition in the ferroelectric phase. These observations also suggest that hybridisation in a traditional $d^0$ picture should have some role in generating the large observed polarisation, above and beyond that produced from purely geometric displacements. A more recent work by \citeasnoun{Tyson2011}, based on the accurate determination of the atomic positions derived from both diffraction and X-ray absorption fine spectra, concurs with the previous experimental works in finding a strongly anisotropic Born effective charge tensor and strong hybridisation effects. 

\subsection{Magnetic transition} \label{subsec-mag}

Hexagonal \emph{R}MnO$_3$ compounds exhibit an antiferromagnetic transition near $T_\mathrm{N}\sim 100$~K due to the superexchange interactions between Mn$^{3+}$ moments. 
In addition, those with magnetic rare-earth ions 
($R$=Ho, Er, and Tm) also show an additional magnetic transition below 10~K, arising from the ordering of the rare-earth moments on the $2a$ Wyckoff sites. The rare-earth moments on the other ($4b$) sites order concurrently with the Mn triangular lattice at $T_\mathrm{N}$, due to a Mn-$R$ superexchange interaction. The rare-earth moments are
thought to align along the $c$ axis and ordered antiferromagnetically within the $ab$-plane~\cite{Alonso2000,Curnoe2006}, although a neutron diffraction study suggested that the
rare-earth moments at the $2a$ site may lie in the $ab$-plane~\cite{Fabreges2008}. In this review, we will focus primarily on the Mn moment ordering.

\subsubsection{Magnetic point groups}

No structural change has been observed at $T_\mathrm{N}$, so the crystallographic space group remains the same as the $P6_3cm$ space group, from which the magnetic space group can be determined. The magnetic
structure was found to have a propagation vector $\mathbf{k} = (0,0,0)$, which gives rise to four possible 1D representations, namely $\Gamma_1$ (A$_1$), $\Gamma_2$ (A$_2$), 
$\Gamma_3$ (B$_1$), $\Gamma_4$ (B$_2$), and two 2D representations, $\Gamma_5$ (A) and $\Gamma_6$ (B), which are illustrated in figure~\ref{fig5}. 
Rather than the $\Gamma$ symbols, the international (Hermann-Mauguin) notation, where symmetry operators that retain time reversal symmetries are primed or underlined,
is also often used in the literature, with the following equivalence: $P6_3cm$ ($\Gamma_1$),
$P6_3\underbar c \underbar m$ ($\Gamma_2$), $P\underbar 6_3c\underbar m$ ($\Gamma_3$), $P\underbar 6_3\underbar cm$ ($\Gamma_4$), $P6_3$ ($\Gamma_5$) and 
$P\underbar 6_3$ ($\Gamma_6$)~\cite{Lorenz2013, Fiebig2003}. The spin arrangements corresponding to these representations are illustrated in figure~\ref{fig5}. 

The magnetic structures represented in figure~\ref{fig5} that preserve the 6-fold rotational symmetry are essentially the 120$^{\circ}$ structure predicted for a classical
Heisenberg antiferromagnet on the triangular lattice, which are either antiferromagnetically ($\Gamma_{1,2,5}$) or ferromagnetically ($\Gamma_{3,4,6}$) coupled along the $c$ axis.
For the $\Gamma_2$ and $\Gamma_3$ representations, the moments can have components along the $c$ axis, which are (anti-)ferromagnetically coupled along the $c$ axis for the 
($\Gamma_3$) $\Gamma_2$ structures. For comparison, the moments are restricted to the hexagonal plane for the $\Gamma_1$ and $\Gamma_4$ structures. In the case of the 1D representations, the in-plane
moments are constrained to be perpendicular ($\Gamma_{1,4}$) or parallel ($\Gamma_{2,3}$) to the $a$ axis, whilst for the 2D representations they may take a constant angle $\phi$
with respects to the crystallographic axis. The 2D representations may also have moment components along the $c$ axis. Finally, $\Gamma_1$ and $\Gamma_3$ are homometric~\cite{Brown2006} so cannot
be distinguished by powder neutron diffraction, as are $\Gamma_2$ and $\Gamma_4$.

\subsubsection{Determination of magnetic structure}

Two main experimental techniques have been used to determine the magnetic structures of $h$-$R$MnO$_3$: neutron diffraction and second harmonic generation (SHG), although magnetometry may
also be used to infer the presence of a $\Gamma_2$ order if a weak ferromagnetic signal is measured, which is not the case for the $h$-$R$MnO$_3$ compounds. Whilst neutron powder diffraction is a common and powerful tool to determine a magnetic structure, it cannot distinguish between the $\Gamma_1$ and $\Gamma_3$ structures, or between the $\Gamma_2$ and $\Gamma_4$ structures. This may be resolved by single crystal polarised neutron diffraction experiments, but the measurements are challenging and have only been reported for HoMnO$_3$ and YMnO$_3$~\cite{Brown2006}.
On the other hand, SHG can, in principle, distinguish between all the possible structures~\cite{Fiebig2000}. For light incident along the $c$ axis, no second harmonic signal
implies either one of  $\Gamma_1$ or $\Gamma_2$ structures, whilst a signal polarised parallel the $a$ axis indicates the $\Gamma_4$ structure and that polarised perpendicular to the $a$ and $c$ axes indicates
the $\Gamma_3$ structure~\cite{Fiebig2003}. Although the $\Gamma_1$ and $\Gamma_2$ structures can be distinguished using light polarised parallel to the $c$ axis, in this case a second harmonic signal from
the ferroelectric polarisation also exists~\cite{Fiebig2005}. Alternatively, the behaviour of the second harmonic signals across a metamagnetic transition under applied
magnetic field can serve to elucidate the zero field magnetic structure~\cite{Fiebig2003}.

\subsubsection{Spin reorientation}

For most $h$-$R$MnO$_3$ compounds, the SHG and neutron data are consistent, yielding a $\Gamma_4$ structure for R = Yb, Tm, and Er in zero field. In the case of YMnO$_3$, powder
neutron diffraction determined the structure to have either the $\Gamma_1$ or $\Gamma_3$ symmetry~\cite{Munoz2000,Lee2008,Sekhar2005, Lee2005}, whilst the SHG work showed a
$\Gamma_3$ structure~\cite{Fiebig2003, Degenhardt2001}. However, a detailed polarised neutron diffraction study~\cite{Brown2006} concluded that it is actually the $\Gamma_6$ structure
(i.e. between $\Gamma_3$ and $\Gamma_4$) but with an angle $\phi=11^{\circ}$, which is closer to the $\Gamma_3$ structure. 

LuMnO$_3$ is another case where the SHG and neutron diffraction disagree, in that SHG found domains with a $\Gamma_4$ structure at high temperatures but $\Gamma_3$ at low
temperatures with an intermediate $\Gamma_6$ phase coexisting with either of the others~\cite{Fiebig2000}. However, neutron diffraction measurements saw no evidence of the $\Gamma_3$ structure at
any temperatures~\cite{Park2010,Tong2012}: some Bragg peaks like 100 expected for the $\Gamma_3$ structureare absent in the experimental data. Furthermore, no evidence of the second phase transition was found in the physical
properties measurements, such as that of dielectric constants~\cite{Katsufuji2001}. 
However, \citeasnoun{Tong2012} reported observing additional peaks in the neutron powder diffraction pattern at low temperatures, and suggest that this arises from an unidentified incommensurate magnetic phase. Whilst this needs to be confirmed independently, it is conceivable that this may explain the SHG measurements. 

The case of HoMnO$_3$ is clearer, however, and a spin reorientation transition from the $\Gamma_4$ to $\Gamma_3$ structures with decreasing temperatures is seen both in
SHG~\cite{Fiebig2003} and neutron~\cite{Vajk2005,Chatterji2014} measurements, as reproduced in figure~\ref{fig6}(a) and (b), respectively. The transition temperature
$T_{\mathrm{SR}}=33$~K is also visible in the physical properties, such as the dielectric constant and magnetic susceptibility~\cite{DelaCruz2005}, heat
capacity~\cite{Lorenz2005} and electric polarisation~\cite{Hur2009}, as shown in figure~\ref{fig7}(a). The mechanism behind this transition is argued to be due to a
change in the sign of the structural trimerisation distortion~\cite{Fabreges2009}, and a spin-lattice coupling via the single-ion anisotropy, which is discussed in detail in section~\ref{subsec-spin-lattice}.

For other $h$-$R$MnO$_3$, although there have been some reports of anomalies in between $T_\mathrm{N}$ and the rare-earth ordering temperature of $\lesssim$10~K in their physical
properties~\cite{Iwata1998,Fan2014}, these observations have not been confirmed by other studies in many cases~\cite{Sugie2002,Katsufuji2002,Sekhar2005}. Furthermore, no
change was observed in the neutron diffraction patterns~\cite{Park2002,Sekhar2005,Fabreges2008,Fabreges2009} or second harmonic generation spectra~\cite{Fiebig2003}. 

Finally, in all $h$-$R$MnO$_3$ a metamagnetic transition occurs under applied magnetic field from the zero field $\Gamma_3$ or $\Gamma_4$ structure to the $\Gamma_2$
structure, as shown in figure~\ref{fig6}(a), and this transition may be hysteretic~\cite{Fiebig2005}. The phase transitions under applied field have been confirmed by some of physical properties measurements~\cite{Sugie2002,Yen2007}, although \citeasnoun{Yen2007} found
no hysteresis in their data. This latter observation was attributed to the weak ferromagnetic moment induced by spin canting that is permitted in the $\Gamma_2$ phase~\cite{Sugie2002}. The combination of this
rare-earth moment together with the sensitivity of the Mn spin direction to the lattice and the Mn-R coupling leads to a very rich magnetic phase diagram for HoMnO$_3$
with intriguing critical behaviour at low temperatures~\cite{Choi2013}.

\subsection{Spin-lattice coupling} \label{subsec-spin-lattice}

The strong nearest neighbour superexchange interaction between the Mn spins favours a structure where the direction of the moments rotates by 120$^{\circ}$ between
neighbours. However, this leaves the spin free to adopt an overall rotation angle $\phi$ with respects to the crystallographic axes. For example, first principles calculations by
\citeasnoun{Solovyev2012} suggested that this direction is set by the single-ion anisotropy, which in turn is determined by the $K_1$ structure distortion that shifts the
Mn ions along the direction of one of the three Mn-O$_{\mathrm{eq}}$ bonds. This \emph{trimerisation} distortion is illustrated in figure~\ref{fig7}. If the Mn ion is shifted
towards the equilateral oxygen (the Mn $x$ coordinate is less than $\frac{1}{3}$, giving small trimers, in figure~\ref{fig7}(b)), then the moments tend to align in this
direction and the magnetic structure is either the $\Gamma_1$ or $\Gamma_4$ structures. On the other hand, if they are shifted away ($x>\frac{1}{3}$, figure~\ref{fig7}(c)), then the moments prefer to
be perpendicular to the bond, giving either the $\Gamma_2$ or $\Gamma_3$ structure~\cite{Solovyev2012}. The interlayer exchange interactions then determine which of these
possible states are adopted. Interestingly, \citeasnoun{Solovyev2012} found that for both YMnO$_3$ ($x>\frac{1}{3}$) and LuMnO$_3$ ($x<\frac{1}{3}$), the interlayer interactions are
antiferromagnetic,
but that in both cases the second neighbour interplanar interaction between overlapping triangles $J_2^{c}$ always has a smaller magnitude than that between neighbouring triangles ${J_2^{c}}'$ 
(as denoted in figure~\ref{fig7}), which thus favours the $\Gamma_3$ (YMnO$_3$) or $\Gamma_4$ (LuMnO$_3$) structures as the $J_2^{c}$ pairs favour a ferromagnetic alignment.

However, the differences in total energy for these structures ($\Gamma_3$ or $\Gamma_4$) due to the single-ion anisotropy is quite small that alternative
calculations by \citeasnoun{Das2014} gives the $\Gamma_3$ structure as the ground state of LuMnO$_3$. Furthermore, it is quite difficult to determine the $x$ coordinate from powder
diffraction measurements so that for YMnO$_3$, which has been well studied, values vary between $x=$~0.3208--0.336 at room temperature~\cite{Munoz2000,Park2010}. For other $h$-$R$MnO$_3$, too, in some cases both $x>\frac{1}{3}$ and $x<\frac{1}{3}$ have been reported for the same compounds, so it
is difficult to establish systematic trends between the crystal and magnetic structures definitively. Nonetheless, the spin reorientation transition observed in HoMnO$_3$
presents a way to test this prediction: above $T_{\mathrm{SR}}$ where the structure is $\Gamma_4$ one should expect to observe $x<\frac{1}{3}$ whilst below
$T_{\mathrm{SR}}$ the structure is $\Gamma_3$, implying $x>\frac{1}{3}$, so that at the transition one expects to see $x=\frac{1}{3}$. Neutron diffraction measurements by
\citeasnoun{Fabreges2009}, reproduced in figure~\ref{fig9}(b), appears to support this hypothesis, albeit with sizeable uncertainties. 

Despite this, the effect of the magnetic ordering on the crystal lattice is clear. Anomalies have been observed in the physical properties at
$T_\mathrm{N}$: in the elastic constants~\cite{Poirier2007} and dielectric permittivity~\cite{Katsufuji2001} as shown in figure~\ref{fig7}(c) and (d). The lattice constants
and unit cell volume have also been observed to deviate from that expected from a Debye-Gr\"uneisen model~\cite{Park2010}, as demonstrated in figure~\ref{fig7}(b).
However, the most striking illustration of this spin-lattice coupling is the astonishing observation by \citeasnoun{Lee2008} of the strong enhancement of the $K_1$
distortion below $T_\mathrm{N}$ in YMnO$_3$ (LuMnO$_3$), where the Mn $x$ coordinate increases (decreases) significantly from $\frac{1}{3}$ below $T_\mathrm{N}$, as reproduced in
figure~\ref{fig9}(a). This may be explained if the gain in the single-ion anisotropy (SIA) energy by further displacing the Mn ions is greater than the costs in the elastic energy.

Another facet of the strong spin-lattice coupling is the observation that the magnetic domains in $h$-$R$MnO$_3$ are clamped to the ferroelectric domains~\cite{Fiebig2002}. 
There are three possible structural rotational directions of the MnO$_5$ polyhedra in the $ab$ plane, denoted $\alpha$, $\beta$ and $\gamma$. The ferroelectric domains 
are then defined by the two possible directions of tilt of the apical oxygen ions, leading to the six possible structural-polarisation domains $\alpha^{\pm}$,
$\beta^{\pm}$ and $\gamma^{\pm}$, which form the characteristic vortex structure observed in microscopy measurements~\cite{Chae2012,Chae2013}. Each of these domains may be
described by a phase angle $\Phi$, which represents the angle to the displaced apical oxygen ions in that domain, and the sequence
$-120^{\circ}$,~$-60^{\circ}$,~$\cdots$,~$+180^{\circ}$ corresponds to $\gamma^+$,~$\beta^-$,~$\alpha^+$,~$\gamma^-$,~$\beta^+$,~$\alpha^-$. It is energetically favourable for
this phase angle to only change by 60$^{\circ}$ between adjacent domains, which thus favours combined antiphase and ferroelectric domain walls, for example from $\alpha^{+}$ to
$\beta^-$ or $\gamma^-$ but not to $\alpha^-$~\cite{Artyukhin2014,Kumagai2013}. As each pair of antiphase domains $\alpha^{\pm}$, $\beta^{\pm}$ and $\gamma^{\pm}$ are
related to a particular magnetic domain due to the preference of the moments to align along or perpendicular to the direction of the Mn displacement as the result of the
trimerisation distortion, this explains why the magnetic domains are locked to the ferroelectric ones~\cite{Artyukhin2014}. We note that purely magnetic domains, where the
moments are rotated by 180$^{\circ}$ across the domain wall, can also exist within a single ferroelectric domain. In sum, the dependence of the magnetic moment on the
unit cell tripling distortions, which drives the ferroelectric order in the $h$-$R$MnO$_3$, provides the mechanism for the magneto-electric coupling in these materials.

\section{Excitations} \label{sec-excit}

As described in the previous sections, one can get great insight into the behaviour of the $h$-$R$MnO$_3$ compounds from their crystal and magnetic structure and how this
changes with temperature or field. However, arguably the ultimate determination of the microscopic Hamiltonian of the system can only be obtained by studying the dynamics
of the atoms (phonons) and magnetic moments (magnons). This will thus provide complementary information to the static behaviour of the structures and also the coupling between the spins and
the lattice, the subject of the preceding sections. Furthermore, the magnetic excitations from the MnO layers, which form a frustrated two dimensional triangular lattice, are themselves of fundamental
interest. In this section, we will review the optical and neutron spectroscopy studies on the excitation spectra of $h$-$R$MnO$_3$ with a particular attention paid to its connection to the structural issue. 
 
\subsection{Phonons} \label{subsec_ph}

Phonons are quantized portions of energies, describing lattice vibration waves. The properties of these waves are described in the reciprocal spaces. In the long wavelength limit, the
possible vibrating modes are determined from the crystal symmetry while phonon energies are sensitive to the interaction strengths between the atoms. Therefore, long-wavelength optical phonons are sensitive to the changes of crystal symmetry and atom positions. The zone center phonon modes in $h$-$R$MnO$_3$ have been studied experimentally (using Raman, THz and
IR spectroscopies) as well as theoretically (using shell model and first principle calculations)~\cite{Iliev1997,Litvinchuk2004,Fukumura2007,Vermette2010,Ghosh2009,Liu2012,Toulouse2014,Vermette2008,
Goian2010,Kadlec2011,Souchkov2003,Kovacs2012,Zaghrioui2008,Basistyy2014,Rushchanskii2012,Varignon2012}. 

In the high temperature paraelectric $P6_3/mmc$ phase, there are altogether 18 phonon modes, of which 5 are Raman active (A$_{1g}$+E$_{1g}$+3E$_{2g}$) and 6 are IR
active (3A$_{2u}$+3E$_{1u}$). \citeasnoun{Fukumura2007} reported measurements of the Raman spectrum up to 1200~K and observed changes around 1000~K, which they attributed
to a transition from $P6_3cm$ to $P6_3/mmc$. This is in contrast to the observed diffraction patterns, which showed that this transition is above 1200~K, as discussed in
section~\ref{subsec-fe}. Moreover, a more detailed study by \citeasnoun{Bouyanfif2015} showed clear evidence of another transition at 1200~K. Thus, we think the four modes
observed by \citeasnoun{Fukumura2007} should be interpreted within the polar $P6_3cm$ symmetry.

In the ferroelectric $P6_3cm$ phase, the unit cell is tripled, resulting in 60 phonon modes at the $\Gamma$ point: among which 38 are Raman active (9A$_1$+14E$_1$+15E$_2$) and 23 are IR active
(9A$_1$+14E$_1$). Early Raman and IR studies on YMnO$_3$ and HoMnO$_3$ identified many of the modes with the A$_1$, E$_1$ and E$_2$ symmetry and compared these with the shell model
calculations~\cite{Iliev1997,Litvinchuk2003}. In most cases, fewer phonon modes were experimentally observed than are allowed by symmetry, which makes it
difficult to match them with the calculated modes. For example, only 8 (9) out of 14 possible E$_1$ (E$_2$) modes and 7 out of 9 possible A$_1$ modes have been observed
for YMnO$_3$ even in the most extensive Raman and IR measurements~\cite{Toulouse2014,Zaghrioui2008}. Although they have been assigned to the nearest energy modes in the
shell model calculations, some ambiguities still remain in all practical likelihood.

A recent IR measurement on LuMnO$_3$, however, may shed light on this problem, finding 13 E$_1$ modes out of 14~\cite{Basistyy2014}. Adopting the highest energy mode
at 644 cm$^{-1}$ found in Raman spectroscopy~\cite{Vermette2010}, the energies of all the possible E$_1$ modes was determined. Moreover, as the mass of Ho is similar to Lu,
it is reasonable to assume that the phonon energies of HoMnO$_3$ is similar to that of LuMnO$_3$. Therefore, we can assign the phonon modes of HoMnO$_3$ to
the nearest phonon modes in LuMnO$_3$, following the analysis used in~\cite{Basistyy2014}. Note that this mode assignment results in higher phonon energies compared to
the shell model calculations, especially for the low energy modes as shown in table~\ref{tab1}. Such discrepancies may possibly be due to oversimplifications in the shell
model calculations.
Indeed, first principle electronic structures calculations of YMnO$_3$ tend to give higher phonon energies for the low energy E$_1$
modes, when compared to those of the shell model calculations~\cite{Rushchanskii2012,Varignon2012}. Thus further theoretical works on the phonon spectra of RMnO$_3$ with heavy
rare-earth elements are required for a more comprehensive understanding of their lattice dynamics. 

\subsection{Magnons} \label{subsec_mag}

Like phonons, magnons are quantized spin waves in magnetically ordered crystals. They are completely described by their dispersion relation $\omega(\mathbf{q})$, where $\mathbf{q}$ is the wave vector. Measurements of this dispersion are sufficient to determine the underlying interactions that governs the spin dynamics, such as exchange interactions and single ion anisotropies. 

\subsubsection{High energy spin dynamics: super-exchange interaction}

The dominant magnetic interaction that determines the 120$^{\circ}$ spin structure is the nearest neighbor superexchange interaction in the triangular Mn-O layer. Several
inelastic neutron scattering experiments have so far reported the magnon dispersion relations for various $h$-$R$MnO$_3$ compounds
~\cite{Sato2003,Vajk2005,Chatterji2007,Lewtas2010,Fabreges2009,Oh2013,Tian2014,Chaix2014}. A simple spin Hamiltonian including only Heisenberg interactions in Mn-O layer is
given by: 
\begin{equation} \label{eq:ex}
\mathcal{H} = J_1 \sum_{\mathbf{intra}} \mathbf{S}_{i} \cdot \mathbf{S}_{j} + J_2 \sum_{\mathbf{inter}} \mathbf{S}_{i} \cdot \mathbf{S}_{j}
\end{equation}
The two different exchange parameters $J_1$ and $J_2$ are due to the Mn trimerization as shown in figure~\ref{fig7}. The magnon spectra can be calculated using Holstein-Primakoff
operators~\cite{Holstein1940} (see also Appendix~\ref{magcalc}). 

The two different values of the exchange interaction are most apparent in the high energy part of the magnon dispersion along the $[h,$ 1$-$2$h,$ $0]$ direction, as shown in the
figure~\ref{fig10}.
If $J_1 \neq J_2$, the triple degeneracy of the magnons at the K point is lifted, resulting in one doubly degenerate mode at high energy and the other at lower energy. When
$|J_2|>|J_1|$, the high energy mode along the M-K direction is almost degenerate, while three different modes are evident for $|J_1|>|J_2|$. Inelastic neutron scattering
studies have reported that a Hamiltonian with $J_1\neq J_2$ is appropriate for YMnO$_3$ and LuMnO$_3$~\cite{Sato2003,Oh2013} while a Hamiltonian with $J_1=J_2$ describes well the measured excitations of HoMnO$_3$~\cite{Vajk2005}. They are consistent with powder
neutron diffraction results, which found that the Mn $x$ coordinate deviates from $\frac{1}{3}$ for YMnO$_3$ and LuMnO$_3$ while it approaches the $\frac{1}{3}$ position for 
HoMnO$_3$ at low temperatures~\cite{Fabreges2009,Lee2008,Park2010}. 
However, theoretical calculations~\cite{Solovyev2012} using the coordinates reported by \citeasnoun{Lee2008} yielded $J_1/J_2\approx$0.8 ($\approx$1.1) for YMnO$_3$ (LuMnO$_3$),
which is quite different from $J_1/J_2\approx$1.5 ($\approx$6) determined from the inelastic neutron scattering experiments. Therefore, it appears that to explain the
large $J_1/J_2$ ratio determined from the experiments on LuMnO$_3$, a much larger shift of the Mn $x$ position is necessary. However, this is unlikely to be the case since the reported changes in the atom positions are already quite
large~\cite{Lee2008}. Thus, the standard interpretation of the magnon spectra reviewed above may need to be revised, and further effects such as magnon-phonon coupling or
magnon-magnon interactions should be taken into account. 
These will be discusses in sections~\ref{subsec-spinphonon} and~\ref{subsec-magnondecay}.

\subsubsection{Low energy spin dynamics: inter-layer coupling and single ion anisotropy}

Although the Hamiltonian above describes the high energy magnon spectra quite well, the inter-layer super exchange interaction and the single ion anisotropy are necessary to
explain the various possible magnetic structures, as discussed in section~\ref{subsec-spin-lattice}. The inter-layer interaction determines the angle between the spins in
alternating triangular layers while the single ion anisotropies fixes the directions of the spins. The final full spin Hamiltonian thus includes four different exchange parameters ($J_1$, $J_2$, $J_1^c$, and $J_2^c$), an easy-plane anisotropy ($D_1$) and easy-axis anisotropy ($D_2$) (see Appendix~\ref{magcalc}). It turns out that the inter-layer interactions and easy-axis anisotropy are over two orders of magnitude smaller than the dominant in-plane
exchange interactions, evidenced by a small dispersion along the $c^*$ direction and a small spin anisotropy gap~\cite{Sato2003,Fabreges2009,Oh2013}. 

However, it is difficult to uniquely determine these parameters from unpolarized inelastic neutron scattering experiments. For example, the change of the sign in $J_1^c$-$J_2^c$
modifies the magnon intensity along the $[h~0~l]$ direction whilst a 90$^{\circ}$ rotation of the easy-axis anisotropy has the exactly same effects. Therefore, the parameter sets
giving the $\Gamma_1$ ($\Gamma_2$) spin configurations and those giving $\Gamma_3$ ($\Gamma_4$) results in the same magnon spectra (see figure~\ref{fig11}). Thus, unpolarised inelastic neutron scattering,
like unpolarised neutron diffraction, cannot distinguish between the two constituents of a homometric pair.  
Nonetheless, the structures determined from the combination of SHG and diffraction measurements can be used to obtain the exact parameter sets from the analysis of inelastic neutron scattering data.

\subsection{Spin-phonon coupling} \label{subsec-spinphonon}

The mechanism underlying the spin-lattice coupling discussed in section~\ref{subsec-spin-lattice} can be investigated further by measuring the changes in the phonon modes
as the antiferromagnetic order develops or by observing the hybridization of magnon and phonon modes. 
Several IR and Raman measurements have shown that many phonon modes shift in energy below $T_\mathrm{N}$
~\cite{Vermette2010,Fukumura2009,Fukumura2007,Ghosh2009,Vermette2008,Litvinchuk2004,Basistyy2014}. 
For example, \citeasnoun{Vermette2010} found that the E$_2$ mode near 250~cm$^{-1}$, reproduced in figure~\ref{fig12}(a), shows a kink at $T_\mathrm{N}$ and hardens below the
temperature. Further IR studies by \citeasnoun{Basistyy2014} on 
$R$=Ho, Er, Tm, Yb, and Lu as reproduced in figure~\ref{fig12}(b) gave similar behaviour for the phonon energies, and reflect the change in the vibrations of the manganese and oxygen ions within
the triangular plane due to the structure distortion that occurs with the onset of the N\'eel order.
 
A related aspect is the hybridization of magnon and acoustic phonon modes, which have been observed by inelastic neutron scattering. 
 \citeasnoun{Petit2007} found that a gap appears in the transverse acoustic phonon mode of YMnO$_3$ below $T_\mathrm{N}$ at around $q_0\approx 0.185$, as shown in figure~\ref{fig13}.
The observed phonon displacement parallel to the $c^*$ axis indicates that the spin couples to the out-of-plane atomic motions. Further polarized
inelastic neutron scattering studies by \citeasnoun{Pailhes2009} showed that the upper split mode has both nuclear and magnetic character, indicating that it is indeed a hybrid mode.
However, only an anti-crossing behaviour was observed at high $|q|$, whilst at low $|q|$ the magnon spectrum showed no gap. This is different from the well studied magnon
phonon hybridization in materials with strong single ion magnetostriction, which shows a gap opening in both the phonon (high $|q|$) and magnon (low $|q|$) dispersions. 
Furthermore, the reciprocal lattice point at $q_{\mathrm{cross}}\approx0.3$, where the magnon and phonon modes cross, does not coincide with the position $q_0$ of the gap. This then 
implies that the magnon-phonon coupling may also have some $q$ dependence in order to explain the experimental data.

There are three main spin-lattice coupling mechanisms that can exist in the $h$-$R$MnO$_3$: single ion magnetostriction~\cite{Vleck1940}, spin current~\cite{katsura2005},
and exchange-striction~\cite{Dharmawardana1970}.
The hardening, below $T_\mathrm{N}$, of the zone center phonon modes that modulates Mn-O-Mn bond lengths and angles has been attributed to the exchange striction model~\cite{Litvinchuk2004,Vermette2010,Basistyy2014},
whilst the spin rotation transitions, as discussed in section~\ref{subsec-spin-lattice}, results from the equilibrium single ion magnetostriction.
However, there is yet no consensus on the origin of the observed magnon-phonon hybridization. For example, it has been attributed by \citeasnoun{Petit2007} to the dynamic single ion
magnetostriction, in which the motions of the atoms modulates the crystal field of the Mn ions that determines the single-ion anisotropy. \citeasnoun{Pailhes2009}, on the other hand,
favours the spin-current mechanism, where it is the Dzyaloshinsky-Moriya (DM) interactions that are modulated.
We also note that the single ion anisotropy in the $h$-$R$MnO$_3$, $D\approx0.3$~ meV~\cite{Sato2003}, is in the same order of magnitude with FeF$_2$,
where strong magnon-phonon hybridization have been observed~\cite{Hutchings1970,lovesey1972}. For comparison, the component of the DM interaction that gives rise to the spin canting is an
order of magnitude smaller than the single ion anisotropy $D$~\cite{Solovyev2012}. On the other hand, there has been no study of the exchange-striction effects on the magnon-phonon coupling in $h$-$R$MnO$_3$. This is
probably because the exchange-striction effects only allow an anharmonic coupling between magnons and phonons in collinear spin systems, and thus has been theoretically neglected. However, a linear coupling is allowed in noncollinear magnets~\cite{Hasegawa2010,Kim2007} and, moreover, the exchange-striction scenario is believed to be the mechanism underlying the electromagnon
observed in orthorhombic RMnO$_3$~\cite{Aguilar2009}. 
Thus the next question to be answered is how much each of these three different mechanisms contribute to the spin-phonon coupling in the $h$-$R$MnO$_3$. 
 
\subsection{Spontaneous magnon decays} \label{subsec-magnondecay}

The magnon spectra have been interpreted within the linear spin wave theory in section \ref{subsec_mag}. In the linear spin wave theory, terms higher than quadratic in $a_i^\dagger$
(creation operator) and $a_i$ (annihilation operator) are neglected. In this case, a magnon is stable with infinite lifetime. The next higher order terms allowed
in collinear magnets are quartic terms giving interactions between magnons, analogous to the Coulomb interaction in electron systems. Although this results in finite
magnon lifetimes at nonzero temperatures, the magnon is a stable quasiparticle at the zero temperature~\cite{Harris1971,Dyson1956,Bayrakci2013}. In magnets with noncollinear spin
structures, however, the next order terms are the cubic terms, which gives an interaction between one and two magnon states that is otherwise forbidden in collinear magnets. This allows the decay
of a magnon into two magnon states that results in finite magnon lifetimes even at the zero temperature~\cite{Chernyshev2006,Chernyshev2009}. This phenomenon is called
'spontaneous magnon decays' and was recently reviewed by \citeasnoun{Zhitomirsky2013}.

One of the simplest systems with a noncollinear spin structure is the two dimensional triangular lattice Heisenberg antiferromagnet (2D TLHA). Therefore, its spectra have been
most studied theoretically amongst noncollinear magnets~\cite{Chernyshev2006,Starykh2006,Zheng2006a,Zheng2006b,Chernyshev2009,Mourigal2013}. However, the experimental
verification of these theoretical predictions is challenging mainly due to the scarcity of (nearly-)ideal 2D TLHA found in nature. For example, dimensional reduction in
Cs$_2$CuCl$_4$~\cite{Coldea2001,Kohno2007} and strong next nearest neighbor interactions in $\alpha$-CaCr$_2$O$_4$~\cite{Toth2012} make their spin excitation spectra quite
different from that predicted for the ideal 2D TLHA. The $h$-$R$MnO$_3$, in contrast, provides a rare opportunity, as their magnon spectra have proven to be very similar to
those of the ideal case~\cite{Chatterji2007,Vajk2005}. 

A recent inelastic neutron scattering study found the clearest evidence of spontaneous magnon decays in LuMnO$_3$~\cite{Oh2013}. For example, the linewidth of the top most magnon mode is significantly
broadened compared to the experimental resolution near $q=(0.5,0.5,0)$, as shown in figure~\ref{fig14}(a). Furthermore, the energy and the $q$ position at this point coincides with
the regions of large two magnon
density of states, as shown in figure~\ref{fig14}(b). Note that a magnon can only decay into two magnon states with the same momentum and energy since the momentum and energy should be preserved during the
decay process~\cite{Zhitomirsky2013}. Therefore, those two-magnon states overlapping with the single magnon dispersion results in many decay channels. Thus, the observed broadening can be
interpreted as the result of a reduced magnon lifetime due to the enhanced decay channels.

\section{Summary and outlook} \label{sec-summary}

Ever since \citeasnoun{Curie1894} conjectured on `the symmetry in physical phenomena, symmetry of an electric field and a magnetic field', it has long been a dream for material scientists to search for this rather unusual class of materials exhibiting the coexistence of magnetism and ferroelectricity in a single compound. Thanks to the extensive volume of works carried out worldwide over the past decade or so,
we have now expanded the list of such materials far beyond the few that were studied in Russia in 1960s~\cite{Astrov1961,Astrov1969,Smolenskii1968}. This experimental renaissance of multiferroic physics seems to give a long overdue justification to the earlier pioneering theoretical works, mainly in the names of two great scientists:  \citeasnoun{Dzyaloshinskii1958} and \citeasnoun{Moriya1960}.

Of such a long list of multiferroic materials, the hexagonal manganites $R$MnO$_3$ and BiFeO$_3$ stand out most for various reasons. In the case of BiFeO$_3$, most of studies were driven by the fact that it is the only compound showing multiferroic behavior at room temperature: all the other multiferroic materials known to date exhibit this unusual ground state only at low temperature~\cite{Park2014}. On the other hand, hexagonal \emph{R}MnO$_3$ has been extensively investigated by using various methods, both experimental and theoretical for its having a two-dimensional triangular lattice. As we have reviewed in this article, it offers a rare yet fascinating playground, where we can explore the combined physics of multiferroic and frustration effects~\cite{Diep2005,Gardner2010,Ramirez1994}, in addition to testing our understanding of two-dimensional triangular antiferromagnetism~\cite{Collins1997}.

First of all, when the antiferromagnetic ground state kicks in at around 80--100~K with the so-called 120$^{\circ}$ coplanar magnetic structure, lowered by a factor of 6 compared to its Curie-Weiss temperature, it gives rise to an extremely large and unusual in-plane deformation of Mn-O layers~\cite{Lee2008,Poirier2007,Souchkov2003,Litvinchuk2004}. When this in-plane deformation occurs, there are subsequent atomic displacements of similar magnitude along the $c$-axis. So not surprisingly, this gigantic spin-lattice coupling induces an extra 0.5~$\mu$C/cm$^2$ of electric polarization, which then provides the necessary coupling among the three otherwise independent degrees of freedom: lattice, spin, and electric polarization~\cite{Lee2005}. At the same time, this unusual spin-lattice coupling is also seen to play a crucial role in suppressing thermal conductivity~\cite{Sharma2004}.

As if this amazing display of a spin-lattice coupling in the structural studies is not enough, yet more surprises come from the studies of the spin dynamics. Its almost ideal triangular lattice and its readily available high-quality single crystals make it a perfect system to explore the spin dynamics of a Heisenberg spin in a triangular lattice. It turns out that the 120$^{\circ}$ coplanar, noncollinear magnetic structure is actually crucial in hosting the hitherto largely ignored effects of magnon-magnon coupling. For example, our detailed studies of spin waves in LuMnO$_3$ unearthed, for the first time, the three key experimental evidence of the magnon-magnon coupling: a roton-like minimum, flat mode, and magnon decay~\cite{Oh2013}. All three of these effects were previously predicted for a triangular magnetic system with noncollinear magnetic ground states~\cite{Zheng2006a,Chernyshev2006,Starykh2006}. Furthermore, we found more recently that there are nontrivial coupling effects of magnon-phonon on the spin dynamics~\cite{Oh2015}. All these works of spin dynamics further illustrate how intimately connected the structural aspect of the $R$MnO$_3$ physics is to their spin dynamics.

In this review, we have examined the structure and spin dynamics of this interesting class of materials. Furthermore, we have also looked at an interesting possibility by using $h$-$R$MnO$_3$ of how we can further deepen our understanding of two-dimensional triangular antiferromagnetism~\cite{Collins1997}, in particular magnon-magnon~\cite{Zhitomirsky2013} or magnon-phonon coupling~\cite{Wang2008,Aguilar2009}.

\ack{This review is the direct result of our extensive works on the hexagonal \emph{R}MnO$_3$ over the years. Therefore, we should acknowledge all the past and present members of the group, who have contributed to our researches on this material directly or indirectly. Moreover, it goes without saying that we have benefited enormously from our extensive network of collaborations. In particular, we should mention few names who made significant contributions to our understanding of the topics summarized in this review: S-W Cheong, Seongsu Lee, Junghwan Park, T. Kamiyama, Y. Noda, A. Pirogov, D. P. Kozlenko, T. G. Perring, \& W. J. L. Buyers. This work was supported by the research programme of Institute for Basic Science (IBS-R009-G1).}

\appendix

\section{Calculation of magnon dispersion relation and dynamical structure factor} \label{magcalc}

\begin{figure} \label{fig5}
\caption{Magnetic structures based on the space group $P6_3cm$ (No. 185). Four possible magnetic point groups in 1D basis vectors and four intermediate ones are shown at the corner and between them, respectively. Reprinted with permission from \citeasnoun{Fiebig2003}.}
\scalebox{.4}{\includegraphics{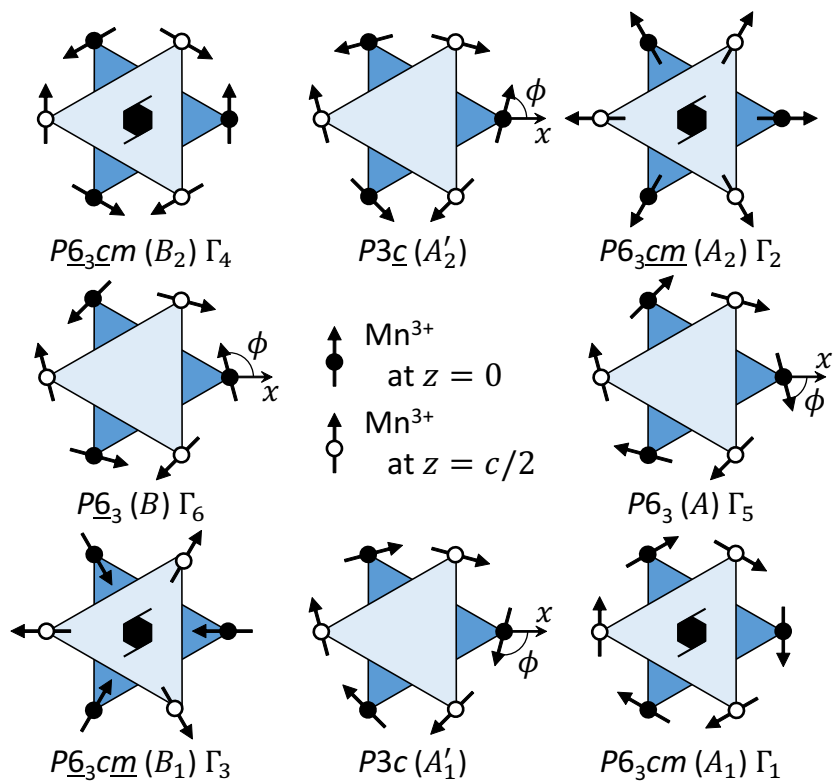}}
\end{figure}

\begin{figure} \label{fig6}
\caption{(a) Phase diagram of several \emph{R}MnO$_3$ using second harmonic generation (SHG) results. (b) Peak intensity of HoMnO$_3$ using powder neutron diffraction data. Both show changes due to the spin reorientation at lower temperatures. Reprinted with permission from \citeasnoun{Fiebig2003} and \citeasnoun{Vajk2005}.}
\scalebox{.6}{\includegraphics{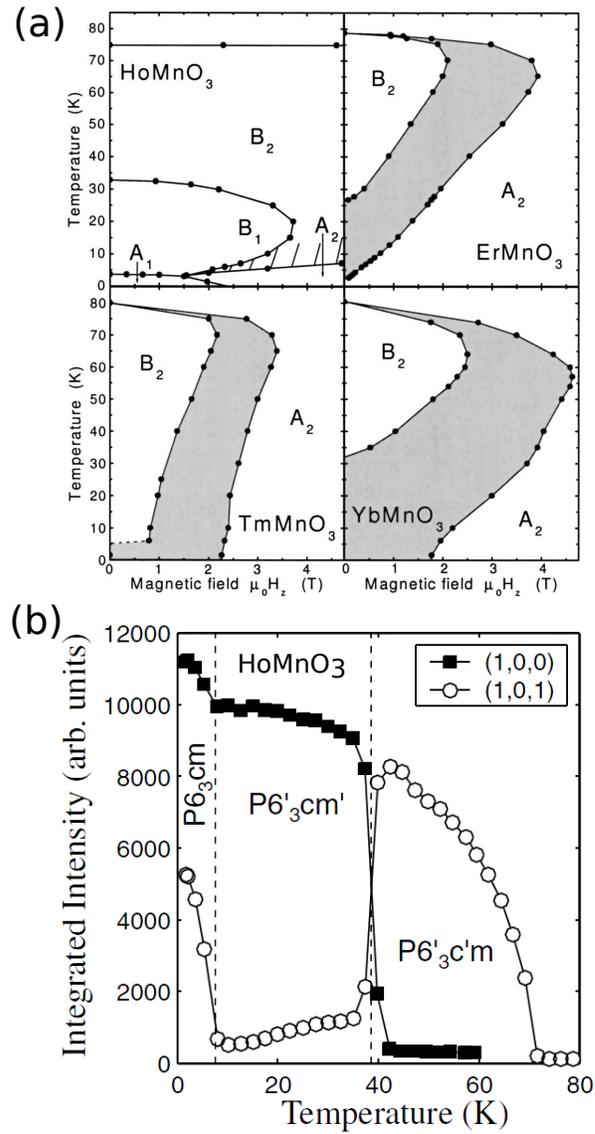}}
\end{figure}

\begin{figure} \label{fig7}
\caption{Several physical properties show distinctive changes at the spin reorientation transition temperature. Temperature dependence is shown of (a) polarization, (b) unit cell volume, (c) dielectric constant and (d) elastic moduli. Reprinted with permission from \citeasnoun{Hur2009}, \citeasnoun{Park2010}, \citeasnoun{Katsufuji2001} and \citeasnoun{Poirier2007}.}
\scalebox{.6}{\includegraphics{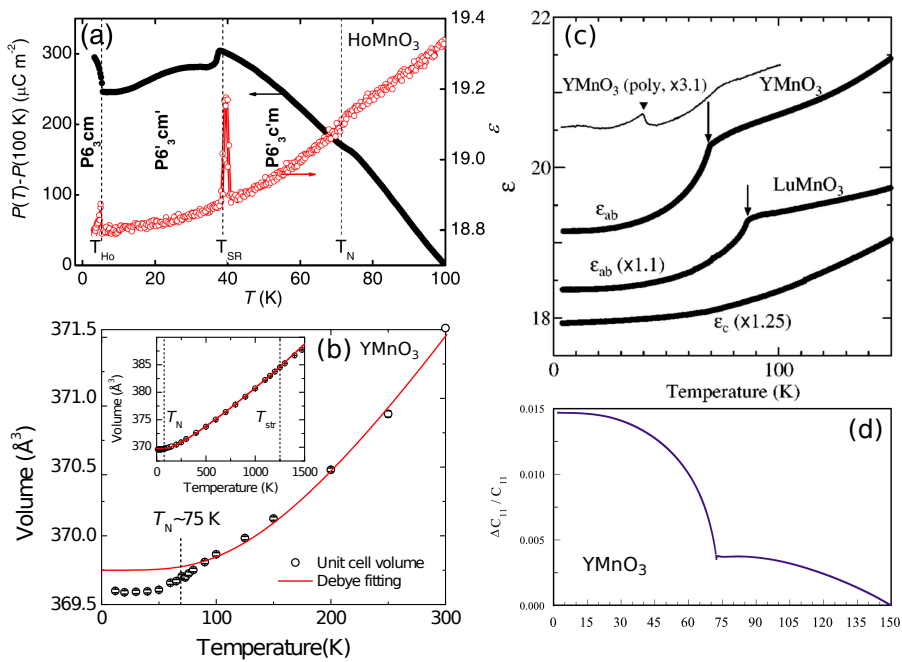}}
\end{figure}

A standard way of calculating magnon spectra for the $\Gamma_4$ spin structure will be covered in this section. The other spin configurations can be handled in a similar manner. The full spin Hamiltonian is given by
\begin{equation} \label{eq:H_full}
\begin{split}
\mathcal{H}  =& J_1 \sum_{\mathbf{intra}} \mathbf{S}_{i} \cdot \mathbf{S}_{j} 
+ J_2 \sum_{\mathbf{inter}} \mathbf{S}_{i} \cdot \mathbf{S}_{j} \\ 
&+ J_1^c \sum_{\mathbf{out intra}} \mathbf{S}_{i} \cdot \mathbf{S}_{j} 
+ J_2^c \sum_{\mathbf{out inter}} \mathbf{S}_{i} \cdot \mathbf{S}_{j} \\ 
&+ D_1 \sum_{i} (\mathbf{S}_{i}^z)^2 
+ D_2 \sum_{i} (\mathbf{S}_{i} \cdot \mathbf{n}_{i})^2 .
\end{split}
\end{equation}
where $\mathbf{n}_i$ is a unit vector parallel to the spin direction at $i$-th site in the $\Gamma_4$ configuration (see figure~\ref{fig5} and \ref{fig7}). The spin operators at six sublattices can be expressed using Holstein-Primakoff operators as shown by the following equations.
\begin{equation} \label{eq:HPop1}
\begin{split} 
S_i^x &= \frac{\sqrt{2S}}{2i}(a_i-a_i^{\dagger})\\
S_i^y &= S-a_i a_i^{\dagger}\\
S_i^z &= \frac{\sqrt{2S}}{2}(a_i+a_i^{\dagger})
\end{split}
\end{equation}
\begin{equation}
\begin{split} 
S_j^x &= -\frac{\sqrt{3}}{2}(S-a_j a_j^{\dagger})-\frac{1}{2}\frac{\sqrt{2S}}{2i}(a_j-a_j^{\dagger})\\
S_j^y &= -\frac{1}{2}(S-a_j a_j^{\dagger})+\frac{\sqrt{3}}{2}\frac{\sqrt{2S}}{2i}(a_j-a_j^{\dagger})\\
S_j^z &= \frac{\sqrt{2S}}{2}(a_j+a_j^{\dagger})
\end{split}
\end{equation}
\begin{equation} \label{eq:HPop2}
\begin{split} 
S_k^x &= \frac{\sqrt{3}}{2}(S-a_k a_k^{\dagger})-\frac{1}{2}\frac{\sqrt{2S}}{2i}(a_k-a_k^{\dagger})\\
S_k^y &= -\frac{1}{2}(S-a_k a_k^{\dagger})-\frac{\sqrt{3}}{2}\frac{\sqrt{2S}}{2i}(a_k-a_k^{\dagger})\\
S_k^z &= \frac{\sqrt{2S}}{2}(a_k+a_k^{\dagger})
\end{split}
\end{equation}
where $i=1, 4$, $j=2, 5$ and $k=3, 6$.
After substituting equation \eqref{eq:HPop1}--\eqref{eq:HPop2} into equation \eqref{eq:H_full} leaving out terms, not higher than quadratic of $a^{\dagger}$ (creation operator) and $a$ (annihilation operator), and
performing Fourier transformation, the Hamiltonian can be rewritten in the following matrix form:
\begin{equation} \label{eq:ham1}
\mathcal{H} = -6 \Delta S(S+1)N + \sum_{k} \mathbf{X}^{\dagger} 
\begin{pmatrix} \mathbf{U} & \mathbf{V} \\ \mathbf{V} & \mathbf{U} \end{pmatrix}
\mathbf{X}
\end{equation}
where
\begin{equation} \label{eq:param1}
\begin{split}
\mathbf{U} &= \begin{pmatrix} 
                        {\mathbf{P}} + \Delta\mathbf{I}_3 & \mathbf{Q} + \mathbf{R}\\ 
                        \mathbf{Q}^{*} + \mathbf{R}^{*}   & {\mathbf{P}}^{*} + \Delta\mathbf{I}_3
                        \end{pmatrix} ,\\
\mathbf{V} &= \begin{pmatrix} 
                        3\mathbf{P}+D_1 \mathbf{I}_3 & 3\mathbf{Q} \\ 
                        3\mathbf{Q}^{*}              & 3\mathbf{P}^{*}+D_1 \mathbf{I}_3 
                        \end{pmatrix} ,
\end{split}
\end{equation}
\begin{equation} \label{eq:param2}
\begin{array}{cc}
\mathbf{X} = \begin{pmatrix} a_{1,k} \\ \vdots \\ a_{6,k} \\ a_{1,-k} \\ \vdots \\ a_{6,-k} \end{pmatrix} , &
\begin{split}
\mathbf{P} &= \begin{pmatrix} 0& A^{*}& C\\A& 0& B^{*}\\ C^{*}& B& 0 \end{pmatrix} ,\\
\mathbf{Q} &= \begin{pmatrix} 0& A'&A'\\A'& 0& A'\\A'& A'& 0 \end{pmatrix} ,\\
\mathbf{R} &= \begin{pmatrix} B'& 0&0\\0& B'& 0\\0& 0& B' \end{pmatrix} ,
\end{split}
\end{array}
\end{equation}
and
\begin{equation} \label{eq:param3}
\begin{split}
A &= \frac{1}{8} \left[J_1+J_2 \left(e^{-ik \cdot b}+e^{-ik \cdot (a+b)} \right) \right] ,\\
B &= \frac{1}{8} \left[J_1+J_2 \left(e^{-ik \cdot a}+e^{ik \cdot b} \right) \right] ,\\
C &= \frac{1}{8} \left[J_1+J_2 \left(e^{ik \cdot a}+e^{ik \cdot (a+b)} \right) \right] ,\\
\Delta &= \frac{1}{2} \left(J_1 + 2 J_2 + 2 J_1^c - 2 J_2^c + D_1 - 2 D_2 \right) ,\\
A' &= \frac{J_1^c}{8} \left(1+e^{-ik \cdot c} \right) ,\\
B' &= \frac{J_2^c}{2} \left(1+e^{-ik \cdot c} \right) .
\end{split}
\end{equation}
Here, $a$ and $b$ denote the lattice unit vectors and $\mathbf{I}_3$ is a 3$\times$3 identity matrix. The numerical diagonalization of the matrix form above results in six magnon modes. The obtained eigenvalues and eigenvectors are used to get the magnon dispersion and dynamical structure factor. For more details of the calculation, see \citeasnoun{White1965} and \citeasnoun{Petit2011}. 

\referencelist[RMnO3_ref]

\onecolumn

\begin{figure} \label{fig8}
\caption{(a) Pattern of Mn trimerization with two different values of the Mn $x$ position with Mn ions forming smaller trimers (red) or larger trimers (blue) on $ab$-plane. The different magnetic exchange interactions are shown for the case of (b) $x<\frac{1}{3}$ and (c) $x>\frac{1}{3}$. }
\scalebox{.4}{\includegraphics{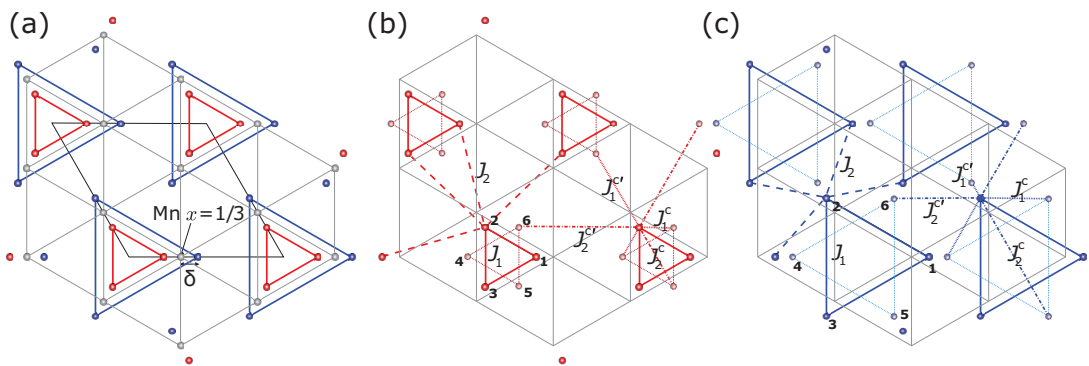}}
\end{figure}

\bgroup
{\setlength{\extrarowheight}{3pt}
\begin{table} \label{tab1}
\caption{Summary of zone center phonon modes calculated from the shell model and measured by Raman and infrared (IR) spectroscopy. Values are taken after the works of [1]~\citeasnoun{Iliev1997}, [2]~\citeasnoun{Toulouse2014}, [3]~\citeasnoun{Zaghrioui2008}, [4]~\citeasnoun{Litvinchuk2004}, [5]~\citeasnoun{Basistyy2014}, and [6]~\citeasnoun{Vermette2010}.}

\begin{center}
\begin{footnotesize}
\renewcommand{\arraystretch}{0.85}
\setlength\tabcolsep{3pt}
\hspace*{-12mm}\begin{tabular}{>{\centering\arraybackslash}p{5cm}|c|rlcc|rlrlc|cc}
                                          &          &    \mcnb{4}{YMnO$_3$}    &       \mcnb{5}{HoMnO$_3$}      &\mct{LuMnO$_3$}\\ \cline{3-13}
Direction and sign of the                 &          & \mct{Shell} &Raman& IR  & \mct{Shell} &\mct{Raman}& IR(TO)&Raman& IR(TO)\\ 
largest displacement                      & Sym.     &\mct{(TO,LO)}&10~K &300~K&\mct{(TO,LO)}&\mct{300~K}& 10~K  & 10~K& 10~K  \\ 
                                          &          &  \mct{[1]}  & [2] & [3] &  \mct{[4]}  & \mct{[4]} & [5]   & [6] & [5]   \\ \cline{1-13}
$+z$(R1), $-z$(R2)                        &   A$_1$  &  147 & 147  & 164 & 154 &  125 & 127  &     &     & 123.5 & 124 &       \\
rot.$x,y$(MnO$_5$)                        &   A$_1$  &  204 & 216  & 211 &     &  195 & 234  &     &     & 223   & 228 &       \\
$+z$(R1,R2), $-z$(Mn)                     &   A$_1$  &  222 & 269  & 262 & 235 &  245 & 270  & 262 & 262 & 256   & 267 &       \\
$x$(Mn), $z$(O3)                          &   A$_1$  &  299 & 301  & 279 & 260 &  291 & 295  & 295 & 295 & 298   & 305 &       \\
$+z$(O3), $-z$(O4), $+x,y$(O2), $-x,y$(O1)&   A$_1$  &  388 & 398  &     & 304 &  404 & 428  & \mct{411} &       &     &       \\
$+z$(O4,O3), $-z$(Mn)                     &   A$_1$  &  423 & 467  & 434 & 432 &  430 & 460  & 427 & 427 &       & 435 &       \\
$+x,y$(O1,O2), $-x,y$(Mn)                 &   A$_1$  &  492 & 496  & 467 & 486 &  468 & 474  & 463 & 463 & 486.1 & 475 &       \\
$+z$(O1,O2), $-z$(Mn)                     &   A$_1$  &  588 & 601  &     & 562 &  598 & 614  &     &     & 580.5 &     &       \\
$+z$(O1), $-z$(O2)                        &   A$_1$  &  662 & 662  & 685 &     &  673 & 673  & 685 & 685 &       & 692 &       \\[8pt]
$+x,y$(Mn,O3,O4), $-x,y$(R1,R2)           &   E$_1$  &  117 & 118  &     &     &  107 & 110  &     &     & 151.5 &     & 147.1 \\
$+x,y$(R1), $-x,y$(R2)                    &   E$_1$  &  147 & 149  &     & 162 &  143 & 143  &     &     &       &     & 155.5 \\
$+x,y$(R2), $-x,y$(R1)                    &   E$_1$  &  158 & 158  &     &     &  149 & 149  &     &     & 165.5 &     & 162.1 \\
$+x,y$(O1,O2), $-x,y$(R1,R2)              &   E$_1$  &  212 & 231  &     & 207 &  231 & 231  &     &     &       &     & 182.4 \\
$x,y$(Mn,O3), $z$(O1,O2)                  &   E$_1$  &  233 & 245  &     & 249 &  247 & 253  &     &     & 245   &     & 270.5 \\
$+x,y$(O1,O2), $-x,y$(O3)                 &   E$_1$  &  250 & 337  &     & 299 &  262 & 336  &     &     & 266.5 &     & 273.5 \\
$+x,y$(O1,O2,O3), $-x,y$(O4,Mn)           &   E$_1$  &  353 & 367  &     & 380 &  337 & 358  &     &     & 292.5 &     & 303.3 \\
$+x,y$(O1), $-x,y$(O2)                    &   E$_1$  &  390 & 403  &     & 400 &  359 & 397  & \mct{354} & 308   &     & 313   \\
$+x,y$(O1), $-x,y$(O2)                    &   E$_1$  &  410 & 415  &     & 416 &  398 & 410  & \mct{369} & 368   & 385 & 368   \\
$+x,y$(O4,O3), $-x,y$(O2,O1,Mn)           &   E$_1$  &  459 & 477  &     &     &  471 & 491  & \mct{419} &       &     & 415   \\
$+x,y$(O4,O3,O1,O2), $-x,y$(Mn)           &   E$_1$  &  492 & 527  &     &     &  497 & 537  & \mct{480} & 420   &     & 428   \\
$x,y$(O4)                                 &   E$_1$  &  559 & 559  &     &     &  568 & 571  &     &     &       &     & 528   \\
$x,y$(O3)                                 &   E$_1$  &  586 & 589  &     & 594 &  585 & 586  &     &     & 591   &     & 600   \\
$x,y$(O3), $-x,y$(O4)                     &   E$_1$  &  635 & 635  &     &     &  648 & 648  & \mct{636} &       & 644 &       \\[8pt]
$x,y$(R1,R2,Mn)                           &   E$_2$  &  \mct{71}   &  85 &     &  \mct{64}   &     &     &       &     &       \\
$+x,y$(Mn,O3,O4), $-x,y$(R1,R2)           &   E$_2$  &  \mct{108}  &     &     &  \mct{96}   &     &     &       &     &       \\
$+x,y$(R1), $-x,y$(R2)                    &   E$_2$  &  \mct{136}  & 142 &     &  \mct{137}  & \mct{136} &       &     &       \\
$+x,y$(R2), $-x,y$(R1)                    &   E$_2$  &  \mct{161}  &     &     &  \mct{152}  &     &     &       &     &       \\
$+x,y$(O2,Mn), $-x,y$(O1,O3)              &   E$_2$  &  \mct{212}  &     &     &  \mct{231}  & \mct{221} &       &     &       \\
$z$(Mn,O2,O1)                             &   E$_2$  &  \mct{241}  & 235 &     &  \mct{254}  &     &     &       &     &       \\
$z$(Mn,O1,O2)                             &   E$_2$  &  \mct{245}  & 249 &     &  \mct{265}  &     &     &       & 260 &       \\
$+z$(O2), $-z$(O1), $x,y$(O4)             &   E$_2$  &  \mct{336}  & 309 &     &  \mct{330}  & \mct{295} &       & 315 &       \\
$+x,y$(O1,O2,O4,O3), $-x,y$(Mn)           &   E$_2$  &  \mct{382}  & 376 &     &  \mct{339}  &     &     &       & 345 &       \\
$+x,y$(O1,O4), $-x,y$(O2,Mn)              &   E$_2$  &  \mct{407}  & 418 &     &  \mct{402}  &     &     &       &     &       \\
$+x,y$(O4), $-x,y$(O1,Mn)                 &   E$_2$  &  \mct{458}  & 442 &     &  \mct{468}  & \mct{442} &       & 463 &       \\
$+x,y$(O4,O3), $+x,y$(O1,O2)              &   E$_2$  &  \mct{515}  &     &     &  \mct{523}  &     &     &       &     &       \\
$x,y$(O4)                                 &   E$_2$  &  \mct{557}  &     &     &  \mct{557}  &     &     &       &     &       \\
$x,y$(O4,O3)                              &   E$_2$  &  \mct{580}  &     &     &  \mct{583}  &     &     &       &     &       \\
$x,y$(O3,O4)                              &   E$_2$  &  \mct{638}  & 637 &     &  \mct{649}  &     &     &       &     &       \\
\end{tabular}
\end{footnotesize}
\end{center}
\end{table}}
\egroup

\begin{figure} \label{fig9}
\caption{Mn $x$ position is one of the most important parameters in understanding the antiferromagnetic ordering and the spin reorientations of \emph{R}MnO$_3$. This figures show the temperature dependence of the Mn atomic position for several RMnO$_3$, reprinted with permission from \citeasnoun{Lee2008} and \citeasnoun{Fabreges2009}.}
\scalebox{.5}{\includegraphics{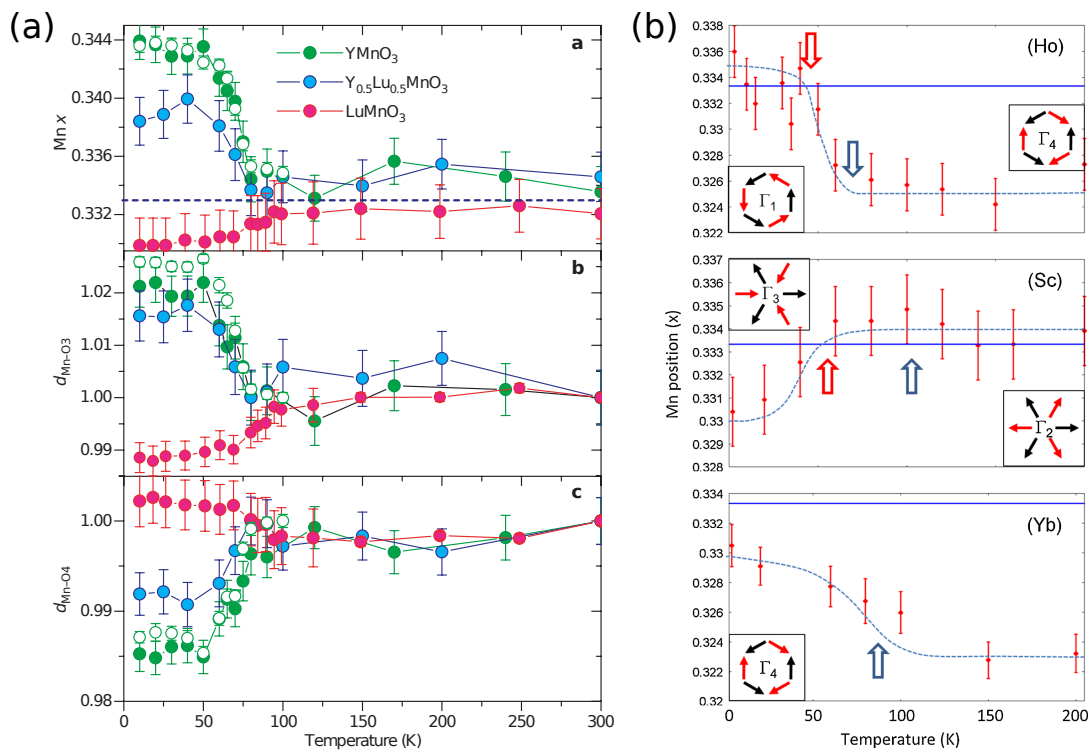}}
\end{figure}

\begin{figure} \label{fig10}
\caption{Magnon spectra for different $J_1$ and $J_2$ values. (a) Experimental magnon dispersion for LuMnO$_3$. (b) Constant-$q$ cut at the high symmetry points. (c, d) Simulation results with $J_1=J_2$ and $J_1<J_2$ models. Reprinted with permission from \citeasnoun{Oh2013}.}
\scalebox{.6}{\includegraphics{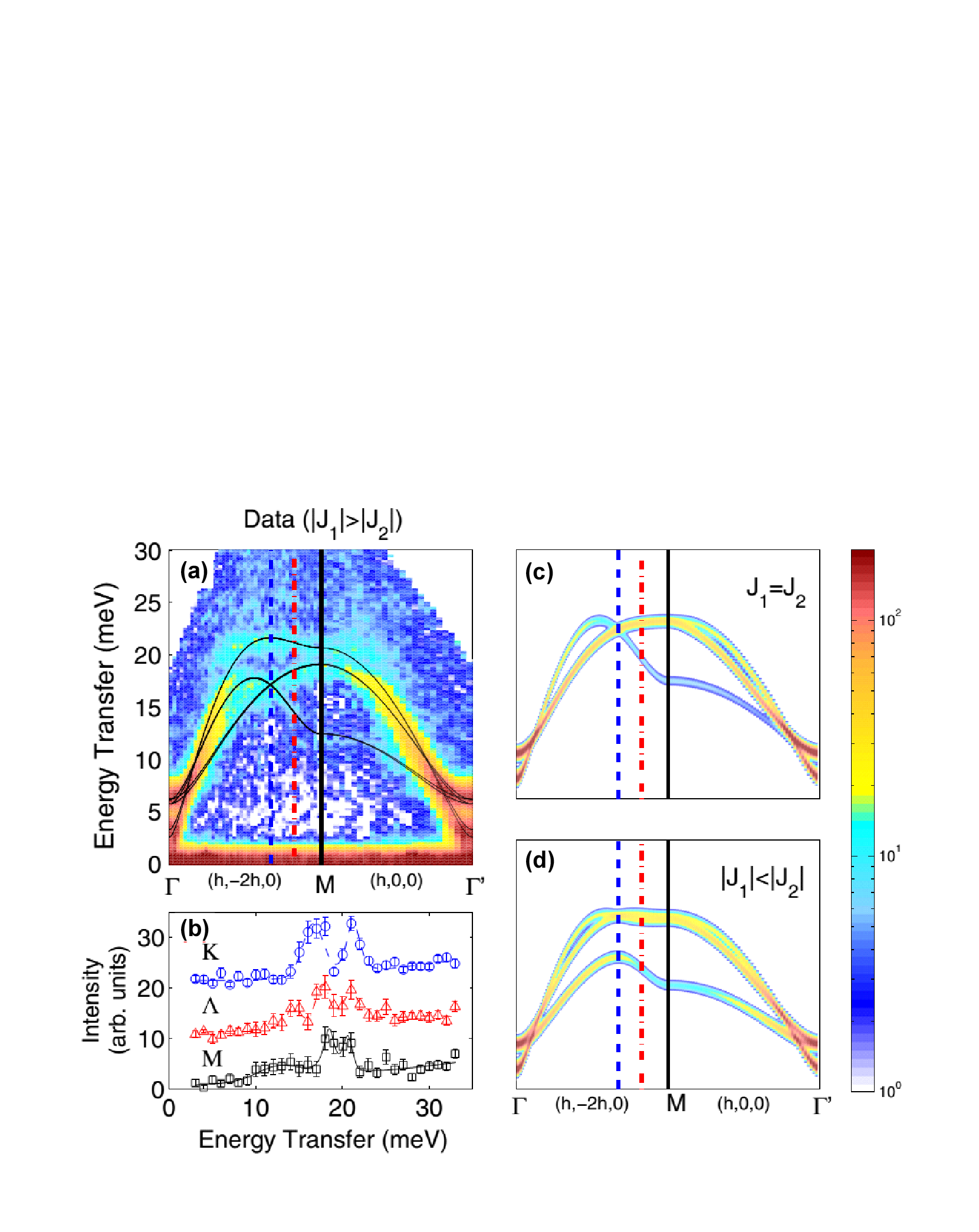}}
\end{figure}

\begin{figure} \label{fig11}
\caption{(Left) Experimental magnon spectra along the $10l$ direction for different compounds: (right) our calculated results that are reproduced by using the following set of parameters: $J_1^c-J_2^c=0.006$, $D_1=0.3$, $D_2=-0.005$~meV for HoMnO$_3$ at 45~K: $J_1^c-J_2^c=0.003$, $D_1=0.3$, $D_2=-0.005$~meV for HoMnO$_3$ at 27~K: $J_1^c-J_2^c=0.015$, $D_1=0.35$, $D_2=-0.045$~meV for YbMnO$_3$: $J_1^c-J_2^c =0.014$, $D_1=0.28$, $D_2=-0.0007$~meV for YMnO$_3$. The experimental results are reprinted with permission from \citeasnoun{Fabreges2009}.}
\scalebox{.7}{\includegraphics{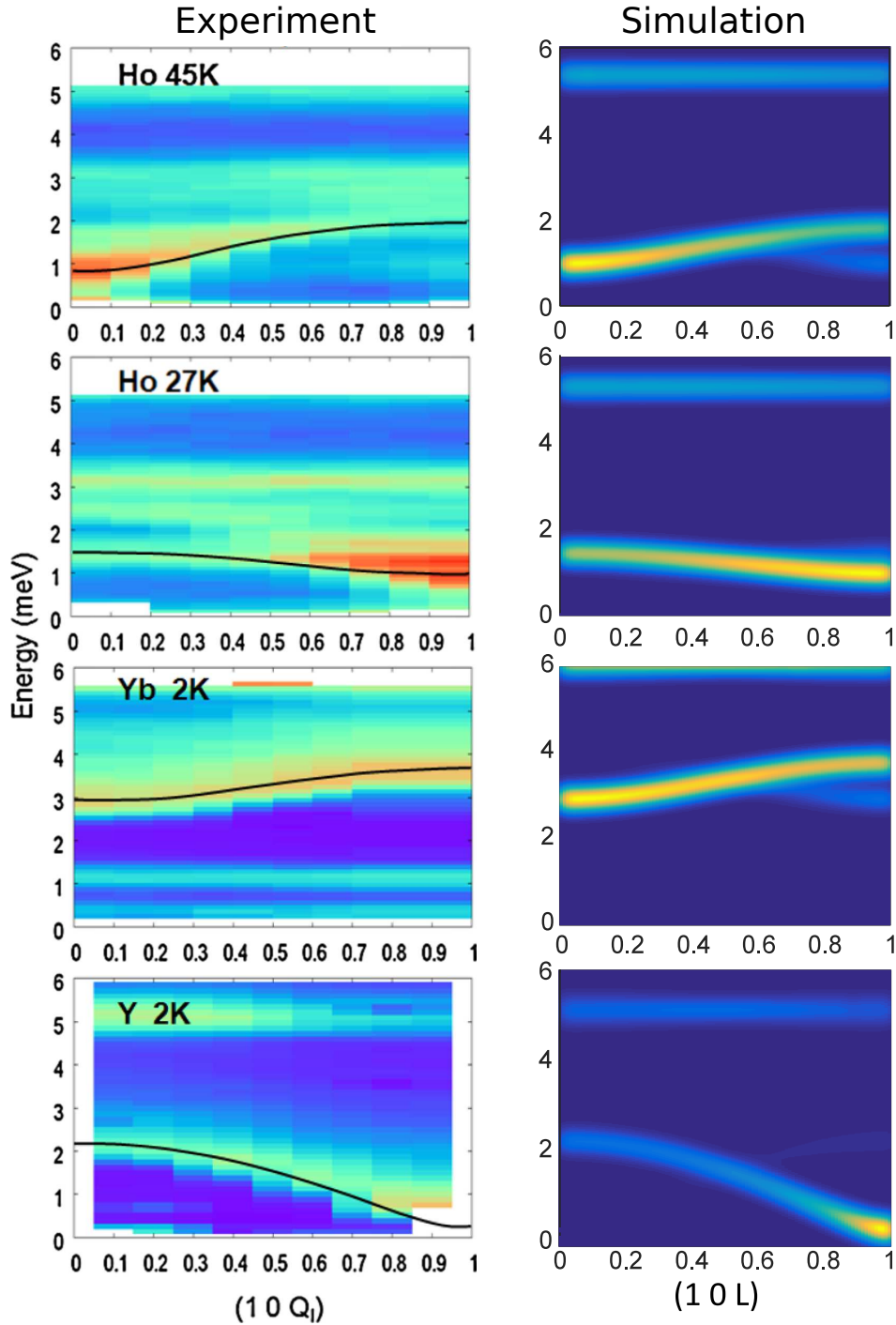}}
\end{figure}

\begin{figure} \label{fig12}
\caption{(a) E$_2$ phonon modes and (b) E$_1$ phonon modes that show visible changes below $T_\mathrm{N}$ for different compounds. Reprinted with permission from \citeasnoun{Vermette2010} and \citeasnoun{Basistyy2014}.}
\scalebox{.6}{\includegraphics{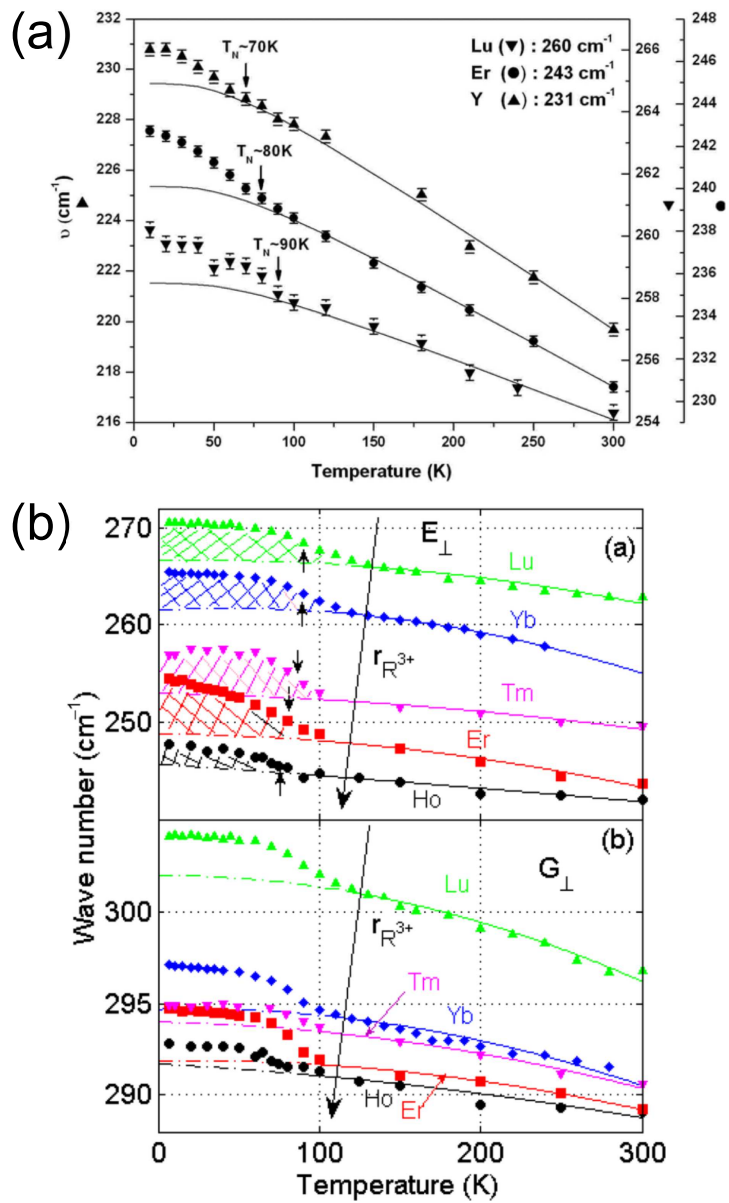}}
\end{figure}

\begin{figure} \label{fig13}
\caption{Experimentally measured and theoretically calculated phonon and magnon modes: (a) a gap in the acoustic phonon mode observed in YMnO$_3$ below $T_\mathrm{N}$ at high $|q|$: (b) no gap in the magnon dispersion at lower $|q|$: (c) the simulation result. Reprinted with permission from \citeasnoun{Petit2007}.}
\scalebox{.7}{\includegraphics{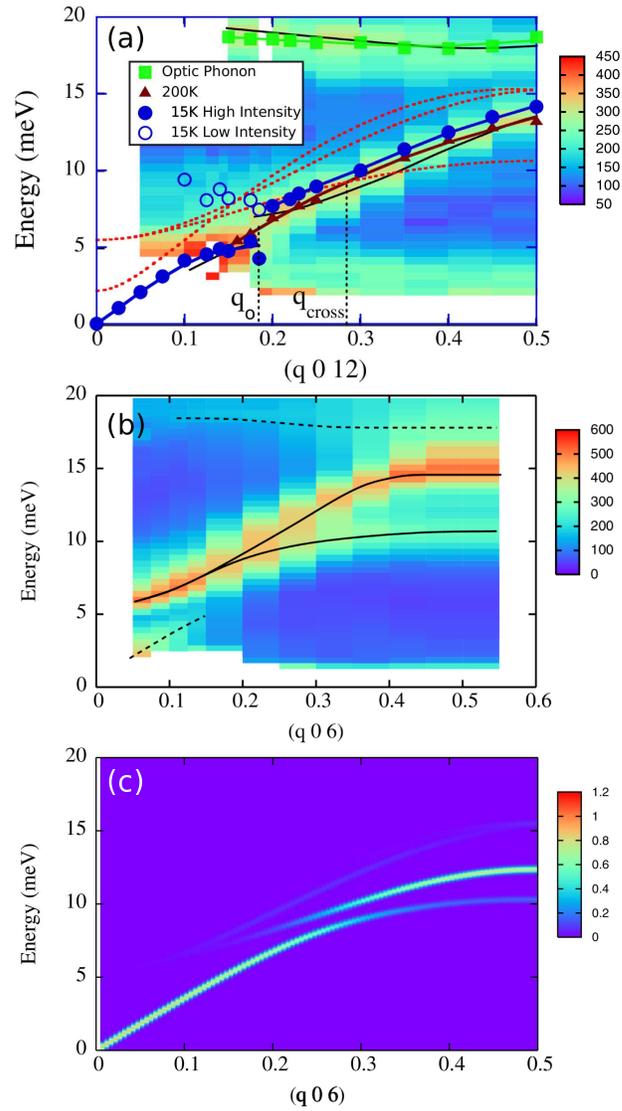}}
\end{figure}

\begin{figure} \label{fig14}
\caption{(a) The linewidth broadening of the top most mode of the spin waves measured at $q=(0.5,0.5,0)$ for LuMnO$_3$ and (b) the calculated two-magnon density of states. Reprinted with permission from \citeasnoun{Oh2013}.}
\scalebox{.5}{\includegraphics{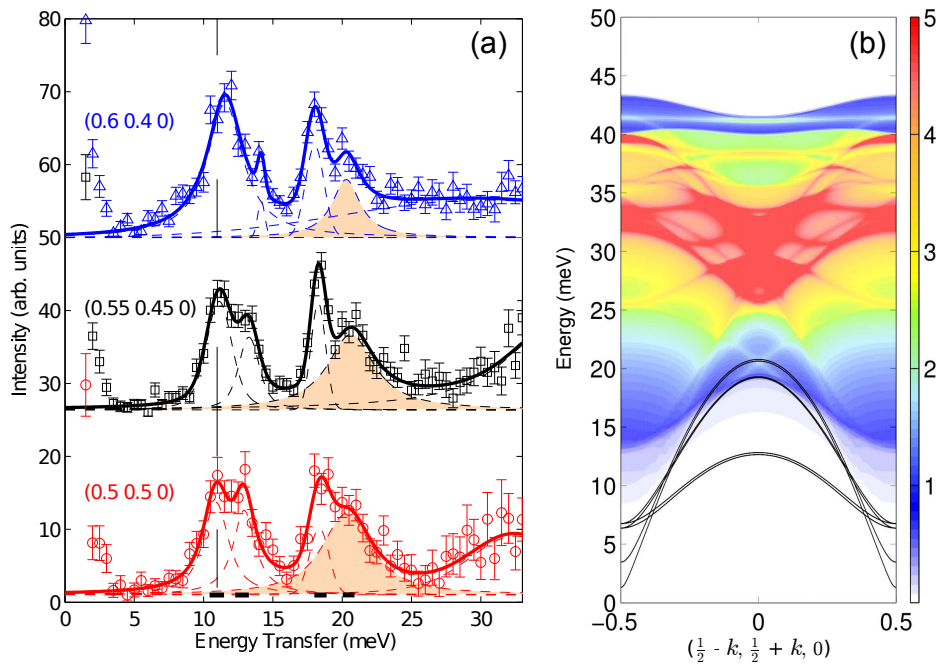}}
\end{figure}

\end{document}